\documentclass[journal=jctcce,manuscript=article]{achemso}

\usepackage[version=3]{mhchem} 
\usepackage{siunitx}
\DeclareSIUnit{\molecule}{molecule}
\usepackage{pgfplots}
\pgfplotsset{compat=1.14}
\usepackage{tikz}
\usepackage{verbatim}
\usepackage{graphicx}
\usepackage{lscape}
\usepackage{booktabs}
\usepackage{longtable}
\usepackage{multirow}


\author{Jacopo Lupi}
\affiliation[SNS]{Scuola Normale Superiore, Piazza dei Cavalieri 7, I-56126 Pisa, Italy}
\author{Cristina Puzzarini}
\affiliation[Unibo]
{Department of Chemistry ``Giacomo Ciamician'', University of Bologna, Via F. Selmi 2, I-40126 Bologna, Italy}
\author{Carlo Cavallotti}
\affiliation[PoliMi]
{Department of Chemistry, Materials, and Chemical Engineering  ``G. Natta'', Politecnico di Milano, I-20131 Milano, Italy}
\author{Vincenzo Barone}
\email{vincenzo.barone@sns.it}
\affiliation[SNS]
{Scuola Normale Superiore, Piazza dei Cavalieri 7, I-56126 Pisa, Italy}

\title[An \textsf{achemso} demo]
  {State-of-the-art quantum chemistry meets variable reaction coordinate transition state theory to solve the puzzling case of the \ce{H2S + Cl} system}

\abbreviations{IR,NMR,UV}

\begin{document}

\begin{tocentry}
\includegraphics{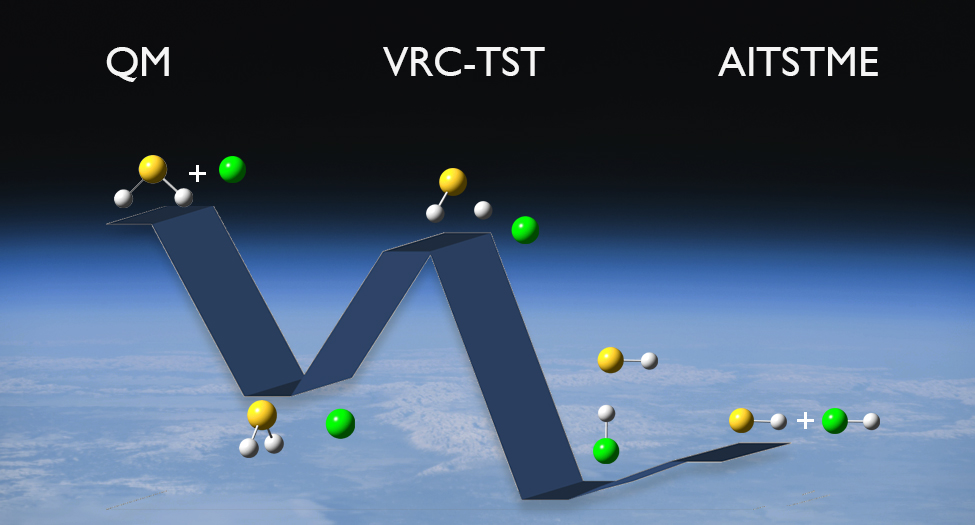}
\end{tocentry}

\begin{abstract}
The atmospheric reaction of \ce{H2S} with Cl has been re-investigated in order to check if, as previously suggested, only explicit dynamical computations can lead to an accurate evaluation of the reaction rate because of strong recrossing effects and the breakdown of the variational extension of transition state theory. For this reason, the corresponding potential energy surface has been thoroughly investigated, thus leading to an accurate characterization of all stationary points, whose energetics has been computed at the state of the art. To this end, coupled-cluster theory including up to quadruple excitations has been employed, together with the extrapolation to the complete basis set limit and also incorporating core-valence correlation, spin-orbit, and scalar relativistic effects as well as diagonal Born-Oppenheimer corrections. This highly accurate composite scheme has also been paralleled by less expensive yet promising computational approaches. Moving to kinetics, variational transition state theory and its variable reaction coordinate extension for barrierless steps have been exploited, thus obtaining a reaction rate constant (\SI{8.16d-11}{\cubic\centi\metre\per\molecule\per\second} at 300 K and 1 atm) in remarkable agreement with the experimental counterpart. Therefore, contrary to previous claims, there is no need to invoke any failure of the transition state theory, provided that sufficiently accurate quantum-chemical computations are performed. The investigation of the puzzling case of the \ce{H2S + Cl} allowed us to present a robust approach for disclosing the thermochemistry and kinetics of reactions of atmospheric and astrophysical interest.
\end{abstract}

\section{Introduction}

Transition state theory (TST) and its variational extension are accepted to be the ``workhorses'' in computational kinetics for several fields ranging from combustion, to atmospheric chemistry, and to astrochemistry (see, e.g., refs. \citenum{C1CP22765C,C7CS00602K,KLIPPENSTEIN2019115}). It is thus fundamental to understand the features of a reactive potential energy surface (PES) that can introduce a breakdown of such theory. In other words, it is important to rationalize its possible limitations. In this respect, some cases of breakdown of TST have been reported in the literature. For instance, Hase and coworkers pointed out that many recrossings can take place in the \ce{Cl- + CH3Cl} gas-phase reaction \cite{ja004077z}, and the same behaviour was observed for the unimolecular isomerizations of NCCN and \ce{CH3CN} \cite{jp045262s}. Another interesting example is offered by the roaming mechanism in \ce{H2CO} photodissociation \cite{B512847C,Townsend1158}. The unsatisfactory results obtained for the \ce{H2S + Cl} reaction by applying canonical variational transition state theory (CVTST) have been interpreted as another failure of TST \cite{resende2008ab}. However, the quantum-chemical approach there employed calls for a deeper re-investigation of the reaction.

To provide a definitive elucidation of the mechanism of the \ce{H2S + Cl} reaction, in the present work, we have investigated its reactive PES by means of state-of-the-art quantum-chemical computations, and coupled them with sophisticated kinetic models still rooted in the TST. In the framework of an ab-initio-transition-state-theory based master equation (AITSTME) treatment, different approaches have been employed to deal with barrierless reactions. The investigation of this specific reaction will lead us to the definition of an accurate protocol for the investigation of the thermochemistry and kinetics of challenging reactions. 

In addition to offer a puzzling case study, the \ce{H2S + Cl} reaction plays an important role in atmospheric chemistry, and might be of relevance also for the investigation of other planetary atmospheres.  
Indeed, the atmospheric sulfur cycle has been the subject of intensive investigations for a long time, mostly because of the need of continuously assessing and monitoring the contribution of anthropogenic sulfur compounds to problems such as acid rain, visibility reduction, and climate modification.\cite{moller1984estimation} In particular, reduced sulfur-containing species are important in the chemistry of the atmosphere; among them, hydrogen sulfide (\ce{H2S}) is one of the simplest, but yet it plays an important role in Earth’s and planetary atmospheres. For what concern the terrestrial environment, its concentration in the atmosphere is significantly due to the decomposition of organic matter and volcanic eruptions, which can inject \ce{H2S} directly into the stratosphere.\cite{seinfeld2016atmospheric,kotra1983chichon} For example, measurements of \ce{H2S} concentrations by UV spectroscopy at volcanic sites in Italy have shown that this quantity can be on the order of hundred parts per million (i.e. much larger than its average atmospheric concentration), \ce{H2S} thus being two to three times more abundant than \ce{SO2}.\cite{o2003real,INGUAGGIATO2018241} However, the anthropogenic emission of \ce{H2S} should not be overlooked.\cite{acp-16-11497-2016,Klimont_2013}

In Earth's atmosphere, \ce{H2S} is mainly removed by the hydroxyl radical (OH) by means of the gas-phase reaction\cite{finlayson1999chemistry}
\begin{equation}
\ce{H2S + OH -> HS + H2O} .
\end{equation}
However, in some marine remote boundary layers and coastal urban areas, the concentration of the chlorine radical (Cl) is larger than that of OH \cite{spicer1998unexpectedly}. Therefore, the reaction of \ce{H2S} with Cl, namely  
\begin{equation}\label{eq:h2s}
\ce{H2S + Cl -> HS + HCl} ,
\end{equation}
is also important. 
Concerning planetary systems, reaction \ref{eq:h2s} can play a role in Venus' atmosphere, the latter being rich of \ce{H2S}.\cite{lewis1970venus,hoffman1980composition,yung1982photochemistry} In addition to its importance in atmospheric processes, reaction \ref{eq:h2s} is of great interest as a prototype for heavy-light-heavy atom reactive systems\cite{polanyi1972concepts}, and --furthermore-- it leads to the production of vibrationally excited HCl molecules, which can be used in infrared chemiluminescence and laser-induced fluorescence studies.\cite{agrawalla1986infrared}

Reaction \ref{eq:h2s} has been studied since 1980, both experimentally and theoretically. Nevertheless, there is still some uncertainty concerning its detailed mechanism. At room temperature, the experimental rate constant spans from \SI{3.7d-11} to \SI{10.5d-11}{\cubic\centi\metre\per\molecule\per\second} \cite{braithwaite1978laser,nesbitt1980laser,clyne1983determination,clyne1984kinetics,nava1985,lu1986,hossenlopp1990kinetics,nicovich1995kinetics,chen2003reaction,gao2015high}.
According to the review study by Atkinson \emph{et al.}\cite{atkinson2004evaluated}, the most reliable results are those by Nicovich \emph{et al.}\cite{nicovich1995kinetics}, who reported an extensive investigation over a wide range of experimental conditions. In that work, the value of the rate constant at room temperature was found to be pressure independent over a wide range, the latter being 33-800 mbar (i.e. 25-600 Torr).

One proposed mechanism is the direct hydrogen abstraction, the reaction thus proceeding through a transition state. Another possibility is offered by an addition/elimination mechanism, which has been discussed by various groups and nowadays seems to be widely accepted. Despite this, to the best of our knowledge, no theoretical work was able to correctly reproduce and interpret the experimental data. Indeed, the computational work by Wilson and Hirst pointed out the existence of a \ce{H2S\bond{...}Cl} adduct, but without being able to connect it with the products, i.e. HS and HCl.\cite{wilson1997ab} In ref.~\citenum{wilson1997ab}, the rate constant for the direct abstraction was computed using conventional transition state theory (cTST), which led to a value, at room temperature, of \SI{2.8d-12}{\cubic\centi\metre\per\molecule\per\second}. However, such rate constant is one order of magnitude smaller than the experimental datum. Resende \emph{at al}. instead investigated the addition/elimination mechanism.\cite{resende2008ab} They obtained a rate constant of \SI{1.2d-9}{\cubic\centi\metre\per\molecule\per\second}, one order of magnitude larger than experimental value, and --as already mentioned above-- they ascribed this discrepancy to a ``breakdown of transition state theory''.

To solve this puzzle, we have undertaken a comprehensive analysis of the whole reaction mechanism using state-of-the-art electronic structure and kinetic models.
The paper is organized as follows. In the next section, the computational methodology is described in some detail, thus introducing the different approaches employed for the electronic structure calculations and kinetics. Then, the results will be reported and discussed: first, the characterization of the reactive PES will be provided, followed by the accurate evaluation of its thermochemistry; then, the reaction rate constants will be addressed. Finally, the major outcomes of this work will be summarized in the concluding remarks.  

\section{Computational Methodology}

In this section, the methodology employed for the characterization of the \ce{H2S + Cl} PES and its energetics will be first of all introduced. Then, we will move to the definition of the models used for the accurate interpretation of the kinetic aspects of the title reaction. 

\subsection{Electronic structure calculations}\label{sect:compmeth}

Several works have shown that double-hybrid functionals in conjunction with basis sets of at least triple-zeta quality represent a remarkable compromise between accuracy and computational cost.\cite{graham2009optimization,sancho2013double,goerigk2014double,mehta2018semi} For the calculation of equilibrium geometries and vibrational frequencies, the B2PLYP functional often approaches, and in some cases even overcomes, the accuracy of the much more computationally expensive CCSD(T) method, when used in conjunction with comparable basis sets (see, e.g., refs. \citenum{Penocchio_rSE_B2PLYP_JCTC15,chemrev.9b00007}). CCSD(T), often denoted as the ``gold standard'' for accurate calculations, stands for the coupled-cluster method including a full account of single and double excitations, CCSD \cite{ManyBody_Methods_ChemPhys}, and a perturbative estimate of triple excitations (CCSD(T)) \cite{Raghavachari-CPL1989_CCSD_T}. In this respect, the recent work by Martin's group has led to the development of the revDSD-PBEP86 functional \cite{santra2019minimally}. This represents a significant improvement with respect to B2PLYP, especially for transition states and non-covalent interactions, also showing very good performances (as B2PLYP) for equilibrium geometries and, especially, vibrational frequencies. Although the very recent D4 model for dispersion contributions \cite{1.5090222} provides some improvement on energy evaluations, the D3(BJ) model \cite{D3,Grimme-JCC2011_DFT-D} is already remarkably accurate. Therefore, since a full analytical implementation of second derivatives of energy is available for the latter \cite{Biczysko-JCTC2010_B2PLYP}, we have decided to rely on the D3(BJ) scheme for incorporating dispersion effects. For all these reasons, in this paper, we have characterized all
stationary points of the reactive PES under consideration with the revDSD-PBEP86-D3(BJ) functional in conjunction with the jun-cc-pV(T+$d$)Z basis set \cite{SummerBasisSets,woon1993gaussian,1.1367373}.

Subsequently, the energetics of all stationary points was accurately determined by exploiting the composite scheme denoted ``HEAT-like'', which is a state-of-the-art approach and will be described in detail in the following. Using this model as reference, the performance of different variants of the so-called ``cheap'' composite scheme\cite{ura,Puzzarini_JPCL2014_GLYdipept} (described in the following as well), shortly denoted as ChS, will be investigated.
For comparison purposes, the CBS-QB3 model \cite{CBS-QB3,CBS-QB3b} will be also considered because it is extensively used in the evaluation of the thermochemistry of reactive systems.

For all levels of theory, the spin-orbit (SO) corrections for the Cl radical as well as for all open-shell species have been computed, within the state-interacting approach implemented in the MOLPRO program \cite{MOLPRO-WIREs,MOLPRO_brief,00268970009483386}, at the complete active space self consistent field (CASSCF) \cite{casscf-1,casscf-2} in conjunction with the aug-cc-pVTZ basis set \cite{Dunning-JCP1989_cc-pVxZ,KDH92} and the full valence as active space. For Cl, calculations of the SO corrections have also been carried out using the multi-reference configuration interaction (MRCI) method \cite{mrci-1,mrci-2,mrci-3} in conjunction with the aug-cc-pV$n$Z sets, with $n$ = T, Q, and 5. The computations for Cl have been performed to calibrate the level of theory to be used for the other radicals. Since it has been noted that the CASSCF values are almost independent of the basis set used (from 274.5 cm$^{-1}$ with aug-cc-pVTZ to 275.7 cm$^{-1}$ with aug-cc-pV5Z) and very close to the results at the MRCI level (276.1 cm$^{-1}$ with aug-cc-pVTZ and 279.5 cm$^{-1}$ with aug-cc-pVQZ), the cheapest level of theory has been chosen for all computations. In passing we note that only at the MRCI/aug-cc-pV5Z level a value of 295.2 cm$^{-1}$, in very good agreement with the experimental result of 293.663 cm$^{-1}$ \cite{Moore}, was obtained. In conclusion, the CASSCF/aug-cc-pVTZ level is expected to provide SO corrections affected by uncertainties not exceeding 0.2 kJ mol$^{-1}$.    

Finally, electronic energies need to be corrected for the zero-point vibrational energy (ZPE) contribution. These corrections have been obtained both within the harmonic approximation and at the anharmonic level. In both cases, they have been computed using the double-hybrid revDSD-PBEP86-D3(BJ) functional in conjunction with jun-cc-pV(T+$d$)Z basis set. Second order vibrational perturbation theory (VPT2)\cite{bloino2012general} has been exploited for the evaluation of anharmonic ZPEs. DFT geometry optimizations and force field computations have been performed with the Gaussian 16 quantum-chemical software \citep{G16C01}.

\subsubsection{Reference structures}

The first issue that needs to be addressed is the effect of the reference geometries on the energetics. Indeed, a reactive PES can be very complicated and characterized by several simultaneous mechanisms. Therefore, the search of the stationary points of a reactive PES and their geometry optimizations might become the computational rate-determining step of the investigation. In this view, it is important to rely on a suitable and reliable level of theory for this purpose. The revDSD-PBEP86-D3(BJ)/jun-cc-pV(T+$d$)Z level seems indeed to meet the requirements. However, further calculations to check the accuracy obtainable in structural determinations and their suitability for energetics have been performed. Concerning the latter point, the ChS approach applied to geometry optimizations has been employed, with the computational details being provided in the specific section. 

Focusing on the reactants well adduct (RW, see Figure~\ref{fig:pes}), as a test case (a radical species with a non-covalent bond), the equilibrium structure has been accurately evaluated by means of a combination of gradient and geometry approaches entirely based on coupled-cluster (CC) theory \cite{ManyBody_Methods_ChemPhys}. First of all, the so-called CCSD(T)/CBS+CV equilibrium structure has been obtained by minimizing the following gradient:
\begin{equation}
\label{eq1} 
\frac{dE_{\mathrm{CBS}}}{dx} =
\frac{dE^{CBS}\mathrm{(HF-SCF)}}{dx} +\frac{d\Delta E^{CBS}\mathrm{(CCSD(T))}}{dx} + \frac{d\Delta E_{CV}}{dx} \; ,
\end{equation}
where the first two terms on the right-hand side are the energy gradients for the extrapolation to the complete basis set (CBS) limit and the last term incorporates the effect of core-valence (CV) correlation.
The exponential formula introduced by Feller \cite{F93} and the two-point $n^{-3}$ expression by Helgaker \textit{et al.} \cite{HKKN97} are used for the extrapolation to the CBS limit of the Hartree-Fock self consistent field (HF-SCF) energy gradient and the CCSD(T) correlation contribution, respectively. The cc-pV$n$Z basis sets \cite{Dunning-JCP1989_cc-pVxZ,woon1993gaussian,v6z-fr,WILSON1996339} have been employed, with $n$=Q, 5 and 6 being chosen for the HF-SCF extrapolation and $n$=Q and 5 for CCSD(T). Since the extrapolation to the CBS limit is performed within the frozen-core (fc) approximation, CV correlation effects have been considered by adding the corresponding correction, $d\Delta E_{CV}/dx$, where the all-electron -- frozen-core energy difference is evaluated employing the cc-pCVQZ basis set \cite{WD95,PD02}. Noted is that the CCSD(T)/CBS+CV equilibrium structure employing the aug-cc-pV$n$Z sets ($n$=T,Q,5 for HF-SCF and $n$=T,Q for CCSD(T)) for the extrapolation to the CBS limit and the cc-pCVTZ basis set for the CV contribution has been obtained for all minima: \ce{H2S}, HS, HCl as well as the reactants-well (RW) and products-well (PW) adducts.    

The contributions due to the full treatment of triple ($\Delta r(\mathrm{fT})$) and quadruple ($\Delta r(\mathrm{fQ})$) excitations have been obtained at the ``geometry'' level, by adding the following differences to the CCSD(T)/CBS+CV geometrical parameters:
\begin{eqnarray}
\label{eq4} 
\Delta r(\mathrm{fT}) & = & r (\mathrm{CCSDT}) - r (\mathrm{CCSD(T)}) \\ \nonumber
\Delta r(\mathrm{fQ}) & = & r (\mathrm{CCSDTQ}) - r (\mathrm{CCSDT}) \; ,
\end{eqnarray}
where $r$ denotes a generic structural parameter. The cc-pVTZ basis set has been used for the fT correction and the cc-pVDZ set for the fQ contribution. This implies that geometry optimizations at the fc-CCSDT\cite{ccsdt1,ccsdt2,ccsdt3}/cc-pVTZ, fc-CCSD(T)/cc-pVTZ, fc-CCSDTQ\cite{ccsdtq}/cc-pVDZ, and fc-CCSDT/cc-pVDZ levels have been performed.

\subsubsection{The HEAT-like approach}

The CC-based approach that has been denoted as HEAT-like takes the HEAT protocol\cite{heat,heat2,heat3} as reference, and starts from the CCSD(T) method. In detail, the scheme (that will be shortly denoted as ``CBS+CV+DBOC+rel+fT+fQ'' in the following) can be summarized as follows:
\begin{equation}
E_{tot}  =  E^{CBS}_{\mathrm{HF-SCF}} + \Delta E^{CBS}_{\mathrm{CCSD(T)}} + \Delta E_{\mathrm{CV}}  + \Delta E_{\mathrm{rel}} + \Delta E_{\mathrm{DBOC}} + \Delta E_{\mathrm{fT}} + \Delta E_{\mathrm{fQ}}  \; .
\label{heat}	   
\end{equation}
In the expression above, ``CBS'' means that CCSD(T) energies --obtained within the fc approximation-- have been extrapolated to the CBS limit. Analogously to what done for the CCSD(T)/CBS+CV gradient scheme, the extrapolation to the CBS limit has been performed in two steps, i.e. the Hartree-Fock self consistent field (HF-SCF) and the CCSD(T) correlation energies have been extrapolated separately. The HF-SCF CBS limit has been evaluated by exploiting the exponential expression introduced by Feller\cite{F93}:
\begin{equation}
E_\mathrm{SCF}(n) = E_\mathrm{SCF}^{CBS} + B \, exp \, (-C \, n) \; .
\label{cbshf}
\end{equation}
For the CCSD(T) correlation contribution, the extrapolation to the CBS limit has been carried out using the $n^{-3}$ formula by Helgaker and coworkers\cite{HKKN97}:
\begin{equation}
\Delta E_\mathrm{corr}(n) = \Delta E_\mathrm{corr}^{CBS} +  A \, n^{-3} \; .
\label{cbscc}
\end{equation}
The cc-pVQZ and cc-pV5Z basis sets have been employed for the CCSD(T) correlation energy, whereas the cc-pV$n$Z sets, with $n$=Q,5,6, have been used for HF-SCF. 

By making use of the additivity approximation, the CV effects have been taken into account by means of the following expression:
\begin{equation}
\Delta E_\mathrm{CV} = E_\mathrm{core+val} - E_\mathrm{val} \; ,
\label{core}
\end{equation}
thus incorporating the CCSD(T) energy difference issuing from all electrons (ae) and fc calculations, both in the cc-pCVQZ basis set\cite{WD95,PD02}.

The diagonal Born-Oppenheimer correction, $\Delta E_{\mathrm{DBOC}}$\cite{dboc1,dboc2,dboc3,dboc4}, and the scalar relativistic contribution to the energy, $\Delta E_{\mathrm{rel}}$, have been computed at the HF-SCF/aug-cc-pVTZ and MP2/unc-cc-pCVQZ (correlating all electrons) levels, respectively, where MP2 stands for M\o ller-Plesset theory to second order\cite{MP34} and ``unc'' denotes the use of the uncontracted basis set. The contributions due to relativistic effects have been evaluated using the lowest-order direct perturbation theory (second-order in 1/c, DPT2)\cite{Kutzelnigg}.

In analogy to equations~\ref{eq4}, corrections due to a full treatment of triples, $\Delta E_{\mathrm{fT}}$, and of quadruples, $\Delta E_{\mathrm{fQ}}$, have computed --within the fc approximation-- as energy differences between CCSDT and CCSD(T) and between CCSDTQ and CCSDT calculations employing the cc-pVTZ and cc-pVDZ basis sets, respectively. 

In addition, to the scheme described above, a variant including the less expensive CCSDT(Q) method\cite{1.1950567,1.2121589,1.2988052} has also been considered:
\begin{equation}
E_{tot}  =  E^{CBS}_{\mathrm{HF-SCF}} + \Delta E^{CBS}_{\mathrm{CCSD(T)}} + \Delta E_{\mathrm{CV}}  + \Delta E_{\mathrm{rel}} + \Delta E_{\mathrm{DBOC}} + \Delta E_{\mathrm{fT}} + \Delta E_{\mathrm{pQ}}  \; ,
\end{equation}
where the only difference with respect to the former approach is the use of CCSDT(Q), which incorporates the quadruple excitations in a perturbative manner, instead of CCSDTQ. This approach will be denoted as ``CBS+CV+DBOC+rel+fT+pQ''.

It should be noted that the HEAT-like schemes also provide all possible intermediate approaches, such as CCSD(T)/CBS, CCSD(T)/CBS+CV and CCSD(T)/CBS+CV+DBOC+rel, together with the results for the single-basis calculations like fc-CCSD(T)/cc-pVQZ and fc-CCSD(T)/cc-pV5Z.

All computations within the HEAT-like schemes have been carried out using the quantum-chemical CFOUR program package \citep{CFour}, with the MRCC code \citep{mrcc} being interfaced to CFOUR in order to perform the calculations including quadruple excitations. 

For all schemes described above, the geometries of the stationary points have been optimized using the double-hybrid revDSD-PBEP86-D3(BJ) in conjunction with the jun-cc-pV(T+$d$)Z basis set. 

\subsubsection{The ChS variants}

The so-called ``cheap'' geometry approach (as already mentioned, shortly denoted as ChS) was initially developed for accurate molecular structure determinations \cite{ura,Thiouracil_B0_PCCP13} and then extended to energetic evaluations \cite{Puzzarini_JPCL2014_GLYdipept} for medium-sized systems. Recently, this scheme has been improved to accurately describe intermolecular, non-covalent interactions (thus leading to the definition of the jun-ChS approach).\cite{jun-chs}
 
While the standard ChS approaches have already been largely applied (the reader is referred to the cited references for more details), the ChS-F12 model has been introduced for the first time. In detail, the general expression of equation \ref{cheap} becomes the following:
\begin{equation}
E_{ChS-F12}=E \mathrm{(CCSD(T)-F12/TZ)} +\Delta E^{CBS} \mathrm{(MP2-F12)} + \Delta E_\mathrm{CV} \mathrm{(MP2-F12)}  \; .	
\label{cheap-f12}
\end{equation}
For evaluating the CC term (using CCSD(T)-F12\cite{ref:f12,werner2011explicitly,knizia2009simplified} within the fc approximation), the cc-pVTZ-F12 basis set\cite{ref:vnzf12} has been used. As in the case of ChS and jun-ChS, the extrapolation to the CBS limit has been performed both in one and two steps. In the case of the two-step procedure, the HF/CBS contribution has been taken from ChS, while for the extrapolation of the correlation contribution with the MP2-F12 method\cite{ref:mp2f12}, two sets of basis sets have been considered\cite{ref:vnzf12}: cc-pVDZ-F12/cc-pVTZ-F12 (extrapolation D,T) and cc-pVTZ-F12/cc-pVQZ-F12 (extrapolation T,Q). For the one-step extrapolation to the CBS limit, the two sets of bases have been used as well. In the case of the D,T extrapolation, the formula \ref{one-s} has been modified accordingly. The CV contribution has been computed using the cc-pCVTZ-F12 basis set.\cite{ref:cvf12} Furthermore, both the F12a and F12b variants\cite{ref:f12} have been employed, and --as described in ref. \citenum{1.3265857}-- the geminal exponent ($\beta$) was set to 1.0 for both the cc-pVTZ-F12 and cc-pVQZ-F12 basis sets.

The final remark concerns the reference geometries used in the energy evaluations. In the case of the standard ChS approaches, the molecular structures employed have been obtained at the same level of theory, the only exception being the tests made on the effect of the reference geometries. This means that an expression analogous to equation \ref{cheap} has been exploited to obtain the geometries of the stationary points. The only difference lies in the fact that, for structural determinations, the extrapolation to the CBS limit has been performed only in one step, according to eq. \ref{one-s}.
Conversely, for ChS-F12, the reference geometries have been determined at the fc-CCSD(T)-F12/cc-pVDZ-F12 level. As mentioned above, to investigate the structural effects on energetics, the jun-ChS model has also been applied to revDSD-PBEP86-D3(BJ)/jun-cc-pV(T+$d$)Z geometries.

For the ChS approaches, the geometry optimizations and energy evaluations have been performed with the MOLPRO quantum-chemistry package\cite{MOLPRO-WIREs,MOLPRO_brief}. 

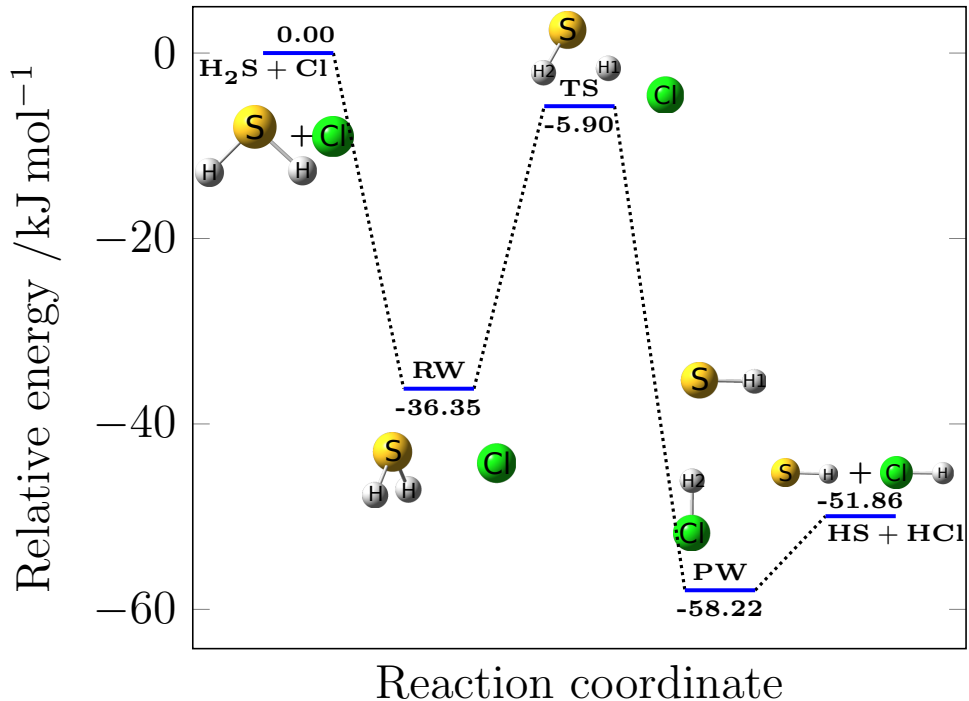
\begin{figure}[ht]
\begin{center}
\begin{tikzpicture}[scale=1.5]
\begin{axis}[clip=false,ylabel=Relative energy /\SI{}{\kilo\joule\per\mol} ,xlabel=Reaction coordinate,xtick=\empty,
legend pos=outer north east,xmin=1,xmax=12,ymax=5,]
]
\node at (axis cs:1.9,-10) {\includegraphics[scale=0.06,angle=0]{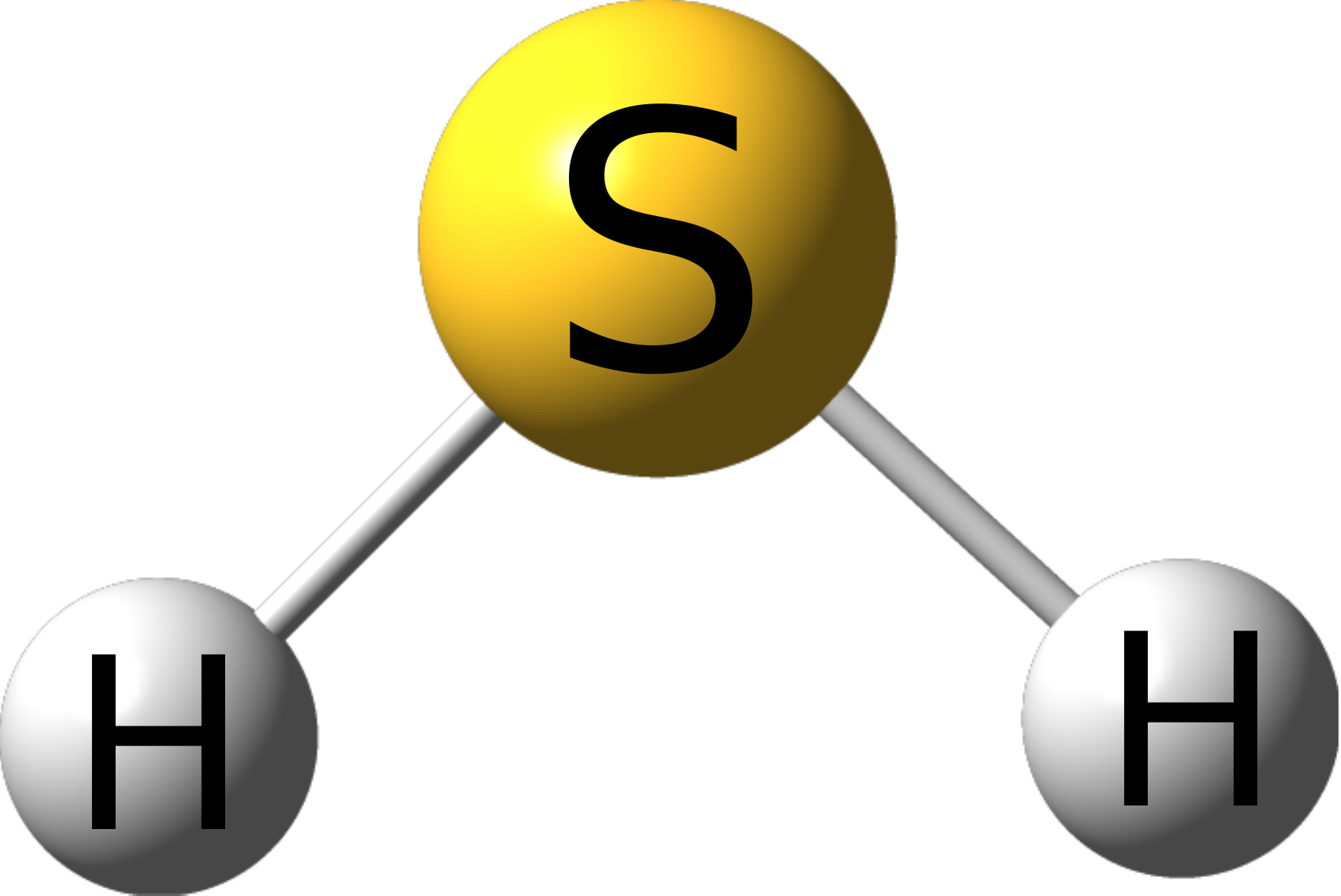}};
\node at (axis cs:3.0,-9) {\includegraphics[scale=0.08,angle=0]{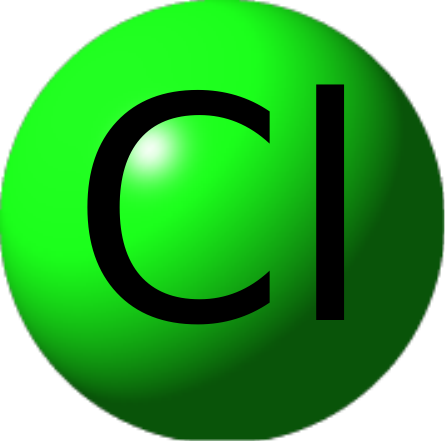}};
\node[font=\tiny] at (axis cs:2,-2) {\bf{\ce{H2S + Cl}}};
\node[font=\tiny] at (axis cs:2.61,2) {\bf{0.00}};
\node[font=\tiny] at (axis cs:2.55,-9) {\bf{+}};
\node at (axis cs:4.5,-45) {\includegraphics[scale=0.18,angle=0]{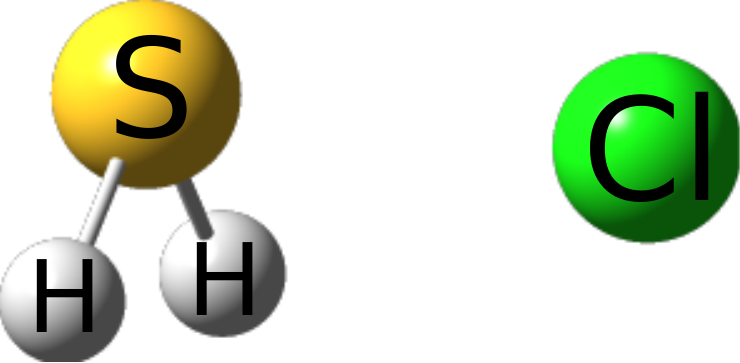}};
\node[font=\tiny] at (axis cs:4.5,-38.20) {\bf{-36.35}};
\node[font=\tiny] at (axis cs:4.5,-34.20) {\bf{RW}};
\node at (axis cs:6.9,-1) {\includegraphics[scale=0.14,angle=0]{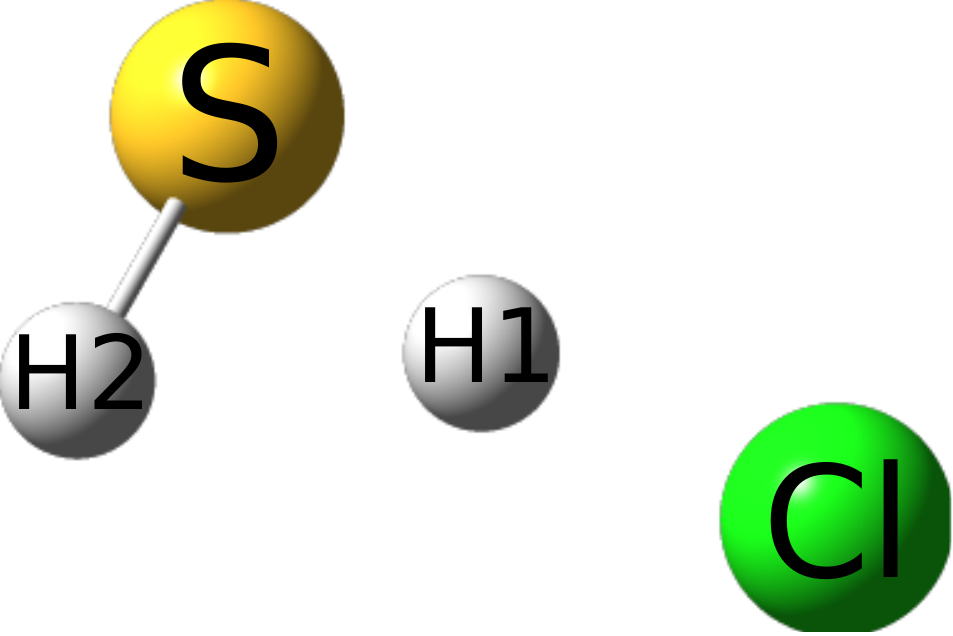}};
\node[font=\tiny] at (axis cs:6.5,-3.73) {\bf{TS}};
\node[font=\tiny] at (axis cs:6.5,-7.73) {\bf{-5.90}};
\node at (axis cs:8.5,-43.51) {\includegraphics[scale=0.17,angle=0]{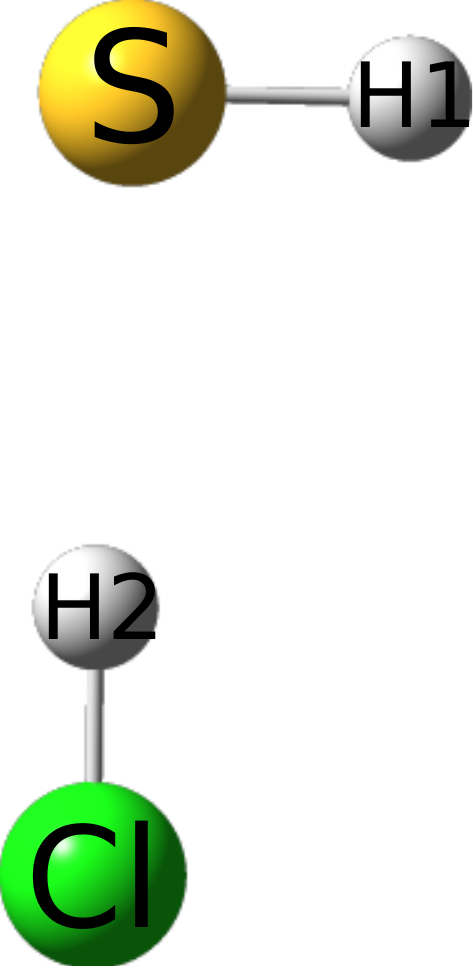}};
\node[font=\tiny] at (axis cs:8.5,-59.94) {\bf{-58.22}};
\node[font=\tiny] at (axis cs:8.5,-55.94) {\bf{PW}};
\node at (axis cs:9.7,-45.25) {\includegraphics[scale=0.08,angle=0]{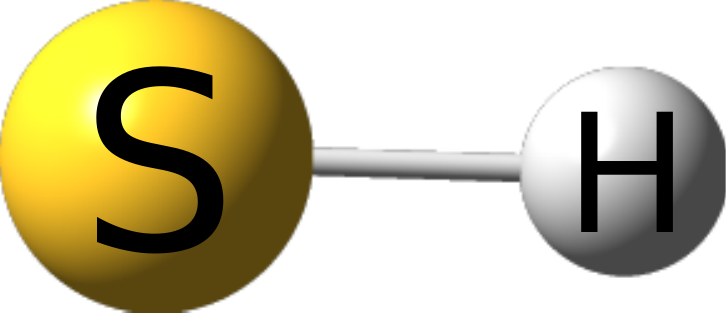}};
\node at (axis cs:11.3,-45.25) {\includegraphics[scale=0.08,angle=0]{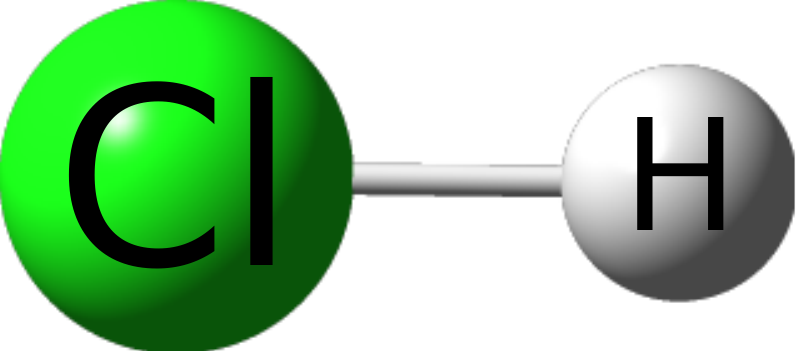}};
\node[font=\tiny] at (axis cs:10.5,-48.20) {\bf{-51.86}};
\node[font=\tiny] at (axis cs:10.5,-45.25) {\bf{+}};
\node[font=\tiny] at (axis cs:11.0,-51.95) {\bf{\ce{HS + HCl}}};
\addplot[color=red,draw=black,line width=0.8pt,densely dotted] coordinates { (3,0)(4,-36.20) };
\addplot[color=red,draw=black,line width=0.8pt,densely dotted] coordinates { (5,-36.20)(6,-5.73) };
\addplot[color=red,draw=black,line width=0.8pt,densely dotted] coordinates { (7,-5.73)(8,-57.94) };
\addplot[color=red,draw=black,line width=0.8pt,densely dotted] coordinates { (9,-57.94)(10,-49.95) };
\addplot[color=red,draw=blue,line width=1pt] coordinates {(2,0) (3,0) };
\addplot[color=red,draw=blue,line width=1pt] coordinates {(4,-36.20) (5,-36.20) };
\addplot[color=red,draw=blue,line width=1pt] coordinates {(6,-5.73) (7,-5.73) };
\addplot[color=red,draw=blue,line width=1pt] coordinates {(8,-57.94) (9,-57.94) };
\addplot[color=red,draw=blue,line width=1pt] coordinates {(10,-49.95) (11,-49.95) };
\end{axis}
\end{tikzpicture}
\caption{Reaction mechanism for the \ce{H2S + Cl} reaction. SO- and ZPE-corrected HEAT-like energies are reported.} \label{fig:pes}
\end{center}
\end{figure}

\subsection{Kinetic Models}\label{sec:kinmod}

The reactive PES for the \ce{H2S + Cl} reaction is summarized in Figure \ref{fig:pes}. As evident from this figure, the Cl atom reacts with \ce{H2S}, thus leading to the \ce{H2S\bond{...}Cl}  intermediate well, denoted as RW. From here onwards, the only available channel is, at least under atmospheric conditions, the addition-elimination reaction: RW isomerizes via the transition state, TS, to the H-bonded \ce{HS\bond{...}HCl} products well, denoted as PW, which then evolves to the products, i.e. HS + HCl. This process can be described in the framework of the AITSTME approach through a three channels, two wells master equation. Depending on the examined temperature and pressure conditions, the global reaction rate is controlled either by the rate of conversion of RW into PW, or by both its rate and that of formation of RW. The rate of decomposition of PW is generally fast with respect to the two other reaction channels and thus does not impact the global reaction rate.

The rate of formation of RW, which proceeds over a barrierless PES, has been determined at three different levels of theory. At the lowest level, the rate constant has been determined using phase space theory (PST)\cite{C7CS00602K}, assuming a $\frac{C}{R^6}$ attractive potential, with the coefficient $C$ obtained by fitting the energies computed at various long-range distances ($R$) of the fragments using the revDSD-PBEP86-D3(BJ)/jun-cc-pV(T+$d$)Z level of theory. Alternatively, the rate constant has been determined using variational transition state theory (VTST) and variable reaction coordinate transition state theory (VRC-TST)\cite{C7CS00602K}, which --among the considered theoretical approaches-- is the most suited to describe properly the large amplitude motions that characterize this reaction channel. 

VTST calculations have been performed within the rigid-rotor harmonic-oscillator (RRHO) approximation over a PES scanned as a function of the distance between Cl and \ce{H2S} in the 3.0-\SI{4.6}{\angstrom} interval using a \SI{0.2}{\angstrom} step. Structures and vibrational frequencies of each point along the PES have been determined at the CASPT2/cc-pVDZ\cite{caspt2_1,caspt2_2,caspt_3} level using a (9e,7o) active space composed of the four \ce{H\bond{-}S} $\sigma$ bonding and antibonding orbitals (4e,4o), and of the three $p$ valence orbitals of chlorine (5e,3o). Higher level energies have been determined at the CASPT2/aug-cc-pVTZ level over a (21e,13o) active space, consisting of the same active space used for the geometry optimization with the addition of the remaining valence electrons of Cl and S and of the three $2p$ orbitals and electrons of S. All CASPT2 calculations have been performed by state averaging over three states using a 0.2 level shift, the only exception being made for high level calculations, which used a 0.25 IPEA shift. The IPEA shift was chosen as it allows to compute with relatively good accuracy the HEAT-like electronic energy of the entrance van der Waals well RW: \SI{-43.7}{\kilo\joule\per\mol} vs \SI{-41.89}{\kilo\joule\per\mol}.

\begin{figure}[ht]
    \centering
    \includegraphics[scale=0.3]{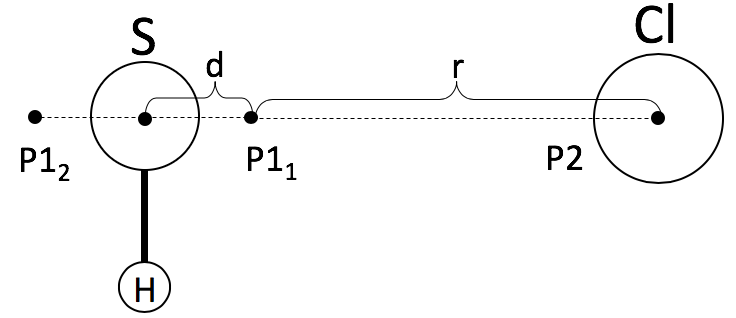}
\caption{Scheme describing the position of the pivot points used to construct the multifaceted dividing surface for VRC-TST calculations. The \ce{H2S} plane is perpendicular to the plane of the picture and the second H atom is hidden behind the S atom.}
    \label{fig:pivot}
\end{figure}

VRC-TST calculations have been performed sampling the PES over multifaceted diving surfaces constructed using three pivot points, positioned as schematized in Figure \ref{fig:pivot}. Two pivot points (P1$_1$ and P1$_2$) were placed in proximity of the S atom, symmetrically displaced along the axis perpendicular to the \ce{H2S} plane and passing from S by a distance $d$, while the third pivot point was centered on Cl.  The multifaceted dividing surface is constructed varying the distance $r$ between Cl and the pivot points, as described by Georgievskii and Klippenstein in ref. \citenum{doi:10.1021/jp034564b}, between 5 and 11 $a_0$ with 1.0 $a_0$ step. Calculations were repeated changing the position of the P1 pivot points varying $d$ between 0.01 and 0.6 $a_0$, using 0.1 $a_0$ steps. Reactive fluxes were computed through Monte Carlo sampling using a 5\% convergence threshold. 

A minimum of 200 sampling points has been taken for each diving surface. The calculated reactive fluxes have then been multiplied by a flat 0.9 factor to correct for recrossing dynamical effects. The 0.9 correction factor comes from the comparison of VRC-TST with trajectory calculations, which showed that VRC-TST total rate coefficients are generally about 10\% greater than those obtained from related trajectory simulations\cite{doi:10.1063/1.1539035}, when VRC-TST is applied at the level of theory used in the present study.\cite{varecof2} Energies have been computed at the CASPT2/cc-pVDZ level using a (5e,3o) active space consisting of the $p$ orbitals of Cl, keeping the structures of \ce{H2S} frozen in its minimum energy configuration and state averaging over three states. The sampled energies have been corrected, using a one dimensional potential function of the distance between the S and Cl atoms, for geometry relaxation, energy accuracy, and SO effects. The correction energy for geometry relaxation ($\Delta E_{\mathrm{GEOM}}$) has been determined at the level of theory used to optimize geometries for VTST calculations, while that for the energy accuracy ($\Delta E_{\mathrm{HL}}$) has been estimated at the higher level used for VTST calculations (CASPT2/aug-cc-pVTZ level over a (21e,13o) active space). Spin-orbit corrections along the PES ($\Delta E_{\mathrm{SO}}$) have been computed using the state interacting method with a Breit-Pauli Hamiltonian and a CASSCF wave function determined with a (9e,7o) active space state and six electronic states. The VRC-TST energy is thus computed as:
\begin{equation}\label{eq:vrctst}
E_{\mathrm{VRC-TST}}=E(\mathrm{CASPT2(5e,3o)/cc-pVDZ})+\Delta E_{\mathrm{GEOM}}+\Delta E_{\mathrm{HL}}+\Delta E_{\mathrm{SO}}.    
\end{equation}

The rate of conversion of RW into PW has been calculated using both conventional and variational TST, as the reaction proceeds through a well defined saddle point. Energies and structures of the stationary points (wells and saddle point) have been evaluated as described in Section \ref{sect:compmeth}. VTST calculations have been performed computing frequencies and structures along a minimum energy path (MEP) determined through intrinsic reaction coordinate (IRC) calculations, at the revDSD-PBEP86-D3(BJ)/jun-cc-pV(T+$d$)Z level, taking 10 steps of 0.03 $a_0$ in both directions. Frequencies along the MEP have been computed both in Cartesian and in internal curvilinear coordinates.\cite{vtst:internal} Tunneling corrections have been taken into account by means of both small curvature theory (SCT)\cite{sct} and the Eckart model\cite{eckart1930penetration}, using an asymmetric potential. 

VRC-TST calculations have been performed using the VaReCoF software\cite{varecof1,varecof2}, while master equation simulations have been carried out using MESS.\cite{georgievskii2013reformulation} VTST calculations (in Cartesian and internal curvilinear coordinates) as well as the determination of stationary points along the MEP to determine the VRC-TST potential and the construction of the input files for VaReCoF computations have been performed using a modified version of EStokTP.\cite{cavallotti2018estoktp} All CASPT2, CASSCF, and SO calculations have been carried out using the MOLPRO program.\cite{MOLPRO_brief} 
For all TST variants, where required, the energies of the stationary points of the PES have been considered at the CBS+CV+DBOC+rel+fT+fQ level, also including SO corrections and anharmonic ZPE contributions.

\section{Results and Discussion}

The reaction mechanism for the \ce{H2S + Cl} reaction is shown in Figure \ref{fig:pes} and, as mentioned in the Introduction, none of the previous computational works provided a complete picture for it. In particular, the presence of both \ce{H2S\bond{...}Cl} and \ce{HS\bond{...}HCl} has been thoroughly investigated only in the present study. In the following, the molecular structures of the stationary points are first presented. Then, the associated thermochemistry and kinetics are reported and discussed.

\subsection{Molecular structures}

\begin{table}
\centering
\caption{Structural parameters of the stationary points of the \ce{H2S + Cl} reaction at different levels of theory. Distances are in \SI{}{\angstrom}, angles in \SI{}{\degree}.}
\label{tab:geom2}
\resizebox{\textwidth}{!}{%
\begin{tabular}{@{}clcccccccc@{}}
\toprule
                                      &       & ChS   & jun-ChS & CC-F12\textsuperscript{\emph{a}}   & DSD-D3\textsuperscript{\emph{b}} & CBS+CV\textsuperscript{\emph{c}} & CBS+CV+fT+fQ\textsuperscript{\emph{d}} & QCISD\textsuperscript{\emph{e}} & \multicolumn{1}{l}{Experiment\textsuperscript{\emph{f}}} \\ \midrule
\multirow{2}{*}{\ce{H2S}}             & r(H-S)                & 1.336 & 1.336 & 1.337  & 1.338 &  1.335 & 1.335 &  -    & 1.3356 \\            
                                      &                       &       &       & (1.338)&       & (1.335)&       &       \\            
                                      & $\theta$(H-S-H)       & 92.2  &  92.2 & 92.2   & 92.5  &  92.3  & 92.3  &  -    & 92.11 \\            
                                      &                       &       &       & (92.3) &       & (92.3) &       &       \\ \midrule   
\multirow{4}{*}{\ce{H2S\bond{...}Cl}} & r(H-S)                & 1.337 & 1.336 & 1.338  & 1.339 &  1.336 & 1.336 & 1.337 \\            
                                      &                       &       &       & (1.339)&       & (1.336)&       &       \\            
                                      & r(S-Cl)               & 2.582 & 2.586 & 2.585  & 2.595 &  2.567 & 2.568 & 2.670 \\            
                                      &                       &       &       & (2.584)&       & (2.585)&       &       \\            
                                      & $\theta$(H-S-H)       & 93.0  & 92.8  & 92.6   & 92.9  &  92.9  & 92.9  & -     \\            
                                      &                       &       &       & (92.8) &       & (92.8) &       &       \\            
                                      & $\theta$(H-S-Cl)      & 87.4  & 87.6  & 87.5   & 88.0  &  87.4  & 87.4  & 88.2  \\            
                                      &                       &       &       & (87.5) &       & (87.5) &       &       \\ \midrule   
\multirow{6}{*}{TS}                   & r(H1-Cl)              & 1.645 & 1.644 & 1.630  & 1.633 &  1.642 & -     & 1.614 \\            
                                      &                       &       &       & (1.642)&       & (1.635)&       &       \\            
                                      & r(H1-S)               & 1.470 & 1.469 & 1.476  & 1.472 &  1.468 & -     & 1.478 \\            
                                      &                       &       &       & (1.471)&       & (1.469)&       &       \\            
                                      & r(H2-S)               & 1.339 & 1.339 & 1.340  & 1.341 &  1.340 & -     & 1.339 \\            
                                      &                       &       &       & (1.341)&       & (1.338)&       &       \\            
                                      & $\theta$(Cl-H1-S)     & 127.0 & 128.2 & 129.8  & 129.6 &  127.2 & -     & 137.0 \\            
                                      &                       &       &       & (128.0)&       & (127.4)&       &       \\            
                                      & $\theta$(H1-S-H2)     & 90.7  & 90.7  & 90.6   & 91.3  &  91.0  & -     & -     \\            
                                      &                       &       &       & (90.8) &       & (90.8) &       &       \\            
                                      & $\varphi$(Cl-H1-S-H2) & 281.0 & 281.0 & 281.4  & 280.4 &  280.9 & -     & -     \\            
                                      &                       &       &       & (281.0)&       & (281.1) &       &       \\ \midrule   
\multirow{5}{*}{\ce{HS\bond{...}HCl}} & r(H2-S)               & 2.508 & 2.506 & 2.484  & 2.481 &  2.492 & 2.492 & -     \\            
                                      &                       &       &       & (2.506)&       & (2.492)&       &       \\            
                                      & r(H1-S)               & 1.341 & 1.341 & 1.343  & 1.343 &  1.341 & 1.341 & -     \\            
                                      &                       &       &       & (1.343)&       & (1.341)&       &       \\            
                                      & r(H2-Cl)              & 1.284 & 1.284 & 1.285  & 1.288 &  1.284 & 1.284 & -     \\            
                                      &                       &       &       & (1.286)&       & (1.284)&       &       \\            
                                      & $\theta$(H1-S-H2)     & 92.4  & 92.3  & 91.8   & 92.6  &  92.0  & 92.0  & -     \\            
                                      &                       &       &       & (91.8) &       & (91.9) &       &       \\            
                                      & $\theta$(Cl-H2-S)     & 176.5 & 176.1 & 175.8  & 176.6 &  175.8 & 175.8 & -     \\            
                                      &                       &       &       & (176.0)&       & (175.8)&       &       \\ \midrule   
\multirow{1}{*}{\ce{HS}}              & r(H-S)                & 1.340 & 1.340 & 1.341  & 1.342 &  1.340 & 1.340 & -     & 1.3406194(3) \\ 
                                      &                       &       &       & (1.342)&       & (1.340)&       &       \\ \midrule   
\multirow{1}{*}{\ce{HCl}}             & r(H-Cl)               & 1.274 & 1.274 & 1.274  & 1.276 &  1.274 & 1.274 & -     & 1.274565598(53) \\            
                                      &                       &       &       & (1.276)&       & (1.274)&       &       \\ \bottomrule

\end{tabular}%
    }
\small
\textsuperscript{\emph{a}} CC-F12 stands for fc-CCSD(T)-F12 in conjunction with the cc-pVDZ-F12 basis set. Values within parentheses have been obtained in conjunction with cc-pVTZ-F12 basis set. \\
\textsuperscript{\emph{b}} DSD stands for revDSD-PBEP86-D3(BJ) in conjunction with the jun-cc-pV(T+$d$)Z basis set. \\
\textsuperscript{\emph{c}} CBS+CV stands for CCSD(T)/CBS+CV, with the aug-cc-pV$n$Z sets ($n$=T,Q) used for the extrapolation to the CBS limit and cc-pCVTZ for the evaluation of the CV contribution. Within parentheses, the results for cc-pVQZ and cc-pV5Z being used for CBS and cc-pCVQZ for CV. See text. \\
\textsuperscript{\emph{d}} CBS+CV+fT+fQ stands for CCSD(T)/CBS+CV augmented for fT and fQ contributions. See text. \\
\textsuperscript{\emph{e}} In conjunction with cc-pV(T+$d$)Z basis set. Values are taken from ref.~\citenum{resende2008ab}. \\
\textsuperscript{\emph{f}} \ce{H2S}: ref.~\citenum{COOK1975237}; HCl: ref.~\citenum{KLAUS1998109}; HS: ref.~\citenum{MARTINDRUMEL20128}.
\end{table}

The structural parameters of the stationary points located on the \ce{H2S + Cl} reactive PES, optimized at different levels of theory, are collected in Table~\ref{tab:geom2}. From the inspection of this table, the first conclusion that can be drawn is that for covalently bonded compounds, i.e. \ce{H2S}, HS and HCl, there is a perfect agreement between ChS, jun-ChS, CCSD(T)/CBS+CV, and revDSD-PBEP86-D3(BJ)/jun-cc-pV(T+$d$)Z results. For the intermediate adducts, i.e. RW and PW, we note again that the revDSD-PBEP86-D3(BJ)/jun-cc-pV(T+$d$)Z covalent bond lengths agree very well with those determined by means of composite schemes. The discrepancies are evident only for the non-covalent distances, i.e. \ce{S\bond{...}Cl} in RW and \ce{S\bond{...}H} in PW. However, the differences are well within 0.01-0.02 \AA, this meaning that their impact on energetics is expected to be negligible (as will be demonstrated in the next section). 

To further test the performance of different ChS approaches and of the revDSD-PBEP86-D3(BJ)/jun-cc-pV(T+$d$)Z level, as mentioned in the computational details section, the CCSD(T)/CBS+CV+fT+fQ scheme has been exploited for the RW adduct. This allowed us to confirm beyond all doubts the accuracy of the ChS models and the suitability of the revDSD-PBEP86-D3(BJ) functional for the characterization of non-covalently bonded systems. Moreover, the results of Table~\ref{tab:geom2} point out the very good performance of the CCSD(T)-F12/cc-pVDZ-F12 level of theory, which is only marginally improved by the use of the cc-pVTZ-F12 basis set. Indeed, also in the case of fc-CCSD(T)-F12 calculations, a very good agreement with the structural parameters issuing from composite approaches can be noted.  

Finally, it is apparent that the results at the QCISD/cc-pV(T+$d$)Z level from ref.~\citenum{resende2008ab} show remarkable deficiencies in the description of the geometrical parameters. In particular, the $\theta$(Cl-H1-S) angle in the transition state is about 10 degrees larger than the corresponding our best-estimated values. Another important deviation is observed for the RW adduct, the \ce{S\bond{...}Cl} distance being too long by $\sim$\SI{0.1}{\angstrom} and the $\theta$(H-S-Cl) too large by $\sim$1 degree.

\subsection{The \ce{H2S + Cl} reaction: Previous works}

In the following, a critical analysis of the previous computational investigations on the \ce{H2S + Cl} reaction is reported.

In the work by Wilson \emph{et al.}\cite{wilson1997ab}, the RW and TS stationary points were characterized by evaluating the electronic energies at the ae-MP4/6-311+G(2df,p) level of theory\cite{krishnan1978approximate,krishnan1980contribution}, using ae-MP2/6-311G** reference structures. Moreover, the addition and abstraction pathways were treated as separate processes instead of combining them in an addition/elimination mechanism, as done in this work. As a consequence, the rate constant was calculated with standard TST considering only the abstraction step. However, a value one order of magnitude lower than the experimental one is mainly explained by an incorrect evaluation of the barrier height.

In a more recent work, in which the PW adduct was not considered as well, Resende \emph{et al.}\cite{resende2008ab} optimized the geometries of the stationary points at the QCISD/cc-pV(T+$d$)Z level of theory.\cite{gauss1988analytical,salter1989analytic} Using these reference structures, the energies were subsequently evaluated at the PMP2 level of theory\cite{PMP21,PMP22,PMP23,PMP24} in conjunction with cc-pV($n$+$d$) basis sets (with $n$ = D, T, Q), extrapolated to the CBS limit using the exponential expression introduced by Feller, and finally added to the corresponding ae-CCSD(T)/cc-pV(T+$d$)Z energies. Resende \emph{et al.} thus employed for the energetics a composite scheme similar to our ChS. Using the aforementioned geometries and energies, the rate constant was subsequently calculated with VTST, thereby obtaining a value that was one order of magnitude higher than the experimental datum. In order to justify their overestimated result, the authors advocated a failure of transition state theory. The role of explicit dynamical effects (e.g. recrossing) was analyzed by performing some explicit trajectory computations. Although the results obtained were not conclusive, they induced the authors to advocate the inadequacy of TST in reproducing the significant role of vibrationally excited \ce{H2S} in stabilizing the \ce{H2S\bond{...}Cl} adduct.

\subsection{The \ce{H2S + Cl} reaction: Thermochemistry}

\begin{table}
\caption{Relative electronic energies$^a$ for the \ce{H2S + Cl} reaction. Values in \SI{}{\kilo\joule\per\mol}.
} 
\label{tab:energie}             
\resizebox{\textwidth}{!}{%
\begin{tabular}{@{}llllll@{}}          
\toprule         
                                              & Reactants   & RW                 & TS           & PW                 & Products      \\ \midrule     
                                              &\ce{H2S + Cl}&\ce{H2S\bond{...}Cl}&              &\ce{HS\bond{...}HCl}& \ce{HS + HCl} \\ \cmidrule(l){2-6}                                    
CCSD(T)/VTZ                                   & 0.00        & -28.29             & 7.74         & -56.10             & -44.90        \\           
CCSD(T)/VQZ                                   & 0.00        & -37.57             & 1.30         & -58.23             & -46.35        \\           
CCSD(T)/V5Z                                   & 0.00        & -41.32             & -0.84        & -59.11             & -46.95        \\           
CCSD(T)/CBS                                   & 0.00        & -44.84             & -3.10        & -60.10             & -47.51        \\           
CBS+CV                                        & 0.00        & -45.09             & -3.42        & -60.30             & -47.65        \\  
CBS+CV+DBOC                                   & 0.00        & -45.09             & -2.05        & -60.15             & -47.64        \\           
CBS+CV+DBOC+rel                               & 0.00        & -45.11             & -2.31        & -59.98             & -47.45        \\           
CBS+CV+DBOC+rel+fT+pQ                         & 0.00        & -45.14             & -3.41        & -60.09             & -47.45        \\     
CBS+CV+DBOC+rel+fT+fQ                         & 0.00        & -45.11             & -3.33        & -60.08             & -47.44        \\    \midrule       
ChS\textsuperscript{\emph{b}}                 & 0.00	    & -42.81	         & -2.49	& -60.29             & -47.94        \\           
                                              &             & [-42.94]           & [-2.54]  & [-60.41]           &  [-47.44] \\ 
jun-ChS\textsuperscript{\emph{b}}             & 0.00	    & -43.89     	 & -3.46        & -60.19            & -47.58        \\           
                                              &   	        &(-43.81)    	 &(-3.45)       &(-60.18)           &(-47.59)       \\  
                                              &             & [-43.89]       & [-3.46]      & [-60.19]          & [-47.58] \\ 
ChS-F12a/F12b CBS(D,T)\textsuperscript{\emph{c}} & 0.00 & -44.64/-44.58 & -1.97/-1.93 & -59.80/-59.80 & -47.11/-47.15   \\          
                                                 &      &[-43.97/-43.91]&[-2.74/-2.70]&[-60.10/-60.10]&[-47.06/-47.09]  \\          
ChS-F12a/F12b CBS(T,Q)\textsuperscript{\emph{c}} & 0.00 & -43.87/-43.81 & -3.32/-3.29 & -60.33/-60.33 &	-47.42/-47.46   \\          
                                                 &      &[-43.54/-43.48]&[-3.39/-3.35]&[-60.27/-60.27]&[-47.28/-47.31]  \\  \midrule
CBS-QB3\textsuperscript{\emph{d}}             & 0.00        & -48.16             & -8.12        & -61.13             & -53.43        \\      \midrule   
revDSD-PBEP86-D3(BJ)\textsuperscript{\emph{e}}& 0.00        & -47.82             & -3.85        & -63.09             & -46.11        \\           
                                              &             &(-45.50)            &(-1.80)       &(-59.43)            &(-46.00)       \\ \midrule                                                                
QCISD/cc-pV(T+$d$)Z\textsuperscript{\emph{f}} & 0.00        & -23.14             & 20.17        & -                  & -43.76        \\           
PMP2/CBS\textsuperscript{\emph{f}}            & 0.00        & -46.69             & -9.08        & -                  & -52.72        \\           
ae-CCSD(T)/CBS\textsuperscript{\emph{f}}      & 0.00        & -46.48             & -7.57        & -                  & -48.16        \\           
\bottomrule                                                                                                                                       
\end{tabular}%
}                                       
\footnotesize	                                
\textsuperscript{\emph{a}} Otherwise stated, references geometries at the revDSD-PBEP86-D3(BJ)/jun-cc-pV(T+$d$)Z level. \\               
\textsuperscript{\emph{b}} Reference geometries at the same level as the energy evaluation. Extrapolation to the CBS limit in one step. Values within parentheses: revDSD-PBEP86-D3(BJ)/jun-cc-pV(T+$d$)Z geometries as reference. Values within square brackets: Extrapolation to the CBS limit in two steps.  \\ 
\textsuperscript{\emph{c}} Reference geometries at the fc-CCSD(T)-F12/cc-pVDZ-F12 level. Extrapolation to the CBS limit in one step. Values within square brackets: Extrapolation to the CBS limit in two steps \\     
\textsuperscript{\emph{d}} Reference geometries at the B3LYP/6-31G(d) level.      \\    
\textsuperscript{\emph{e}} Values obtained in conjunction with the jun-cc-pV(Q+$d$)Z basis set. Within parentheses: results for the jun-cc-pV(T+$d$)Z basis set. In both cases: reference structures at the revDSD-PBEP86-D3(BJ)/jun-cc-pV(T+$d$)Z level. \\
\textsuperscript{\emph{f}} Results from ref. \citenum{resende2008ab}.      
\end{table}

The relative electronic energies, obtained at various computational levels, are detailed in Table \ref{tab:energie}. In particular, fc-CCSD(T) results in conjunction with basis sets of increasing size up to the CBS limit are collected together with the CBS+CV, CBS+CV+DBOC, CBS+CV+DBOC+rel, and CBS+CV+DBOC+rel+fT+fQ values. These allow us to inspect the trend of the relative energies as a function of the basis set as well as the role of the various contributions. 

For all stationary points, it is noted that even for a basis set as large as cc-pV5Z the results are not yet converged, with values being quantitatively accurate only at the CBS limit. A particular remark is warranted for the fc-CCSD(T)/cc-pVTZ level because it is often used in the investigation of reactive PESs, in particular for astrophysical purposes. This level predicts the RW adduct to lie $\sim$28 \SI{}{\kilo\joule\per\mol} below the reactants, which means less stable by about 17 \SI{}{\kilo\joule\per\mol} with respect to what evaluated by more accurate calculations. Furthermore, at this level, the transition state is predicted to emerge above the reactants by more than 7 \SI{}{\kilo\joule\per\mol}. By enlarging the basis set, we note that TS is still emerged with the quadruple-zeta set and becomes barely submerged only when the quintuple-zeta basis is used.   

Moving from the fc-CCSD(T)/CBS level on, it is observed that the CV corrections are small, these being --on average-- of the order of 0.2 \SI{}{\kilo\joule\per\mol}. Roughly of the same order of magnitude are the combined DBOC and scalar relativistic contributions, with the only exception being the transition state, for which the correction amounts to about 1 \SI{}{\kilo\joule\per\mol}. Analogous is the situation for the fT and fQ contributions, with the corresponding correction for TS being, however, in the opposite direction. The overall conclusion is that the CBS+CV level provides results in very good agreement with the CBS+CV+DBOC+rel+fT+fQ model. Furthermore, it is noted that when basis sets more suitable for describing the third-row elements are used (i.e. cc-pV($n+d$)Z), for the CBS+CV model smaller basis sets can be employed. In another test, the extrapolation to the CBS limit has been applied to ae-CCSD(T)/cc-pCV$n$Z ($n$=Q,5) energies. Since the differences with the CBS+CV level lie within 0.2 \SI{}{\kilo\joule\per\mol}, the validity of the additivity approximation for the CV contribution has been confirmed. From Table \ref{tab:energie}, it is also evident that the perturbative (instead of full) treatment of quadruple excitations marginally affects the relative energies. 

Moving to less expensive approaches, the very good performance of all ChS variants deserves to be highlighted. Indeed, ChS, jun-ChS, and ChS-F12 provide very similar results, which deviate --in the wrong cases-- by about 2 \SI{}{\kilo\joule\per\mol} from our best estimates (i.e. CBS+CV+DBOC+rel+fT+fQ). The jun-ChS values obtained using also the revDSD-PBEP86-D3(BJ)/jun-cc-pV(T+$d$)Z reference geometries are reported, within parentheses, in Table \ref{tab:energie}. A perfect agreement between the two sets of jun-ChS relative energies is observed, the differences being well below 0.1 \SI{}{\kilo\joule\per\mol}. Such a comparison confirms the accuracy of the revDSD-PBEP86-D3(BJ)/jun-cc-pV(T+$d$)Z structures for thermochemical studies. In addition, the revDSD-PBEP86-D3(BJ) functional provides accurate results also for the energetics, as clear from the results collected in Table \ref{tab:energie}. We note that this functional performs better when combined with the jun-cc-pV(T+$d$)Z basis set for all minima, while the jun-cc-pV(Q+$d$)Z set seems to be required for correctly evaluating the relative energy of the transition state. 

In Table \ref{tab:energie}, for all ChS variants, relative energies based on extrapolations to the CBS limit performed both in one and two steps are reported. It is worth noting that the two approaches provide very similar results. This is an important outcome because the extrapolation in one step avoids the problems related to basis sets of quintuple-zeta quality for which convergence of the HF-SCF energy can be troublesome, especially for open-shell species. Furthermore, the computation of the corresponding integrals might become particularly expensive for large systems. 

As mentioned in the computational methodology section, both F12a and F12b variants have been employed in the ChS-F12 model. From the results of Table \ref{tab:energie}, it is evident that the two approximations provide nearly coincident results. For this scheme, the extrapolation to the CBS limit using double- and triple-zeta basis sets has also been tested. It is apparent that even in this case the results are very good, showing deviations from the best estimated values well within 2 \SI{}{\kilo\joule\per\mol}. However, this ChS variant is the only one presenting some difference (about 1 \SI{}{\kilo\joule\per\mol}) between the one- and two-step procedures for the extrapolation to the CBS limit.  

As already pointed out in the literature (see, e.g., refs. \citenum{Vazart_JCTC2016_Formamide,D0CP00561D}), despite its widespread use, the CBS-QB3 model provides disappointing results for energetics and, in particular, for barrier heights. In fact, the transition state lies too low in energy by about 5 \SI{}{\kilo\joule\per\mol} with respect to the best estimate.

Finally, the comparison of our results with those from ref. \citenum{resende2008ab} is deserved. First, it is noted that the QCISD/cc-pV(T+$d$)Z level, whose unsuitability in the determination of reference structures has been previously pointed out, provides unreliable energetics, the transition state being predicted $\sim$20 \SI{}{\kilo\joule\per\mol} above the reactants. The situation improves moving to the PMP2 level and, in particular, to the level denoted in Table \ref{tab:energie} as ae-CCSD(T)/CBS, which corresponds, as previously described, to a composite scheme similar to the ChS model. Despite this improvement, the barrier height is underestimated by about 3 \SI{}{\kilo\joule\per\mol}.        

To complete the thermochemistry of the \ce{H2S + Cl} reaction, the relative energies need to be corrected for SO coupling (which was neglected in ref. \citenum{resende2008ab}). Doing so for our best level of theory leads to the values collected in the first row of Table \ref{tab:energiesocasscf}. The major consequence is that the transition state is no longer submerged and lies above the reactants by 0.12 \SI{}{\kilo\joule\per\mol}. However, a further correction should be taken into consideration. Once ZPE is incorporated, the transition state is again submerged by about 5 \SI{}{\kilo\joule\per\mol}. Furthermore, it has to be noted that ZPE corrections evaluated within the harmonic approximation are in very good agreement with the anharmonic values.

The computed standard formation enthalpies (298.15 K, 1 atm) of \ce{H2S} and products (HS and HCl), obtained from our electronic structure calculations and the experimental formation enthalpies of the H, S, Cl atomic species\cite{codata}, are in remarkable agreement with the most accurate experimental data taken from ref. \citenum{codata}, except for that of HS \cite{hsnew}, namely: 141.86 vs. 141.87, -91.82 vs. -92.17 and -20.34 vs. -20.50 \SI{}{\kilo\joule\per\mol} for HS, HCl and H2S, respectively. All details are provided in the SI (see Table S1). From these results, we obtain a sub-\SI{}{\kilo\joule\per\mol} accuracy also for the reaction enthalpy of the title reaction (50.92 vs. 51.24 \SI{}{\kilo\joule\per\mol}).

\begin{table}[ht!]                                      
\caption{Best-estimated relative electronic energies (including spin-orbit) together with ZPE and thermochemical corrections. Values in \SI{}{\kilo\joule\per\mol}. }       
\label{tab:energiesocasscf}                                           
\resizebox{\textwidth}{!}{%
\begin{tabular}{@{}llllll@{}}                                         
\toprule                                                                  
                                    & Reactants    & RW                 & TS    & PW                 & Products    \\ \midrule 
                                    & \ce{H2S + Cl}&\ce{H2S\bond{...}Cl}&       &\ce{HS\bond{...}HCl}&\ce{HS + HCl}\\ \cmidrule(l){2-6}    
CBS+CV+DBOC+rel+fT+fQ\textsuperscript{\emph{a}}& 0.00  & -41.89 & -0.05  & -56.79 & -46.25 \\     
                                               &(3.28) & (0.06) &(0.01) & (0.00) &(2.10) \\            
anharm-ZPE\textsuperscript{\emph{b}}           & 0.00  & 5.54   & -5.85 & -1.43  & -5.61  \\            
harm-ZPE\textsuperscript{\emph{b}}             & 0.00  & 5.86   & -5.27 & -1.55  & -5.86  \\            
$\Delta H^\circ - \Delta H_\circ^\circ$ \textsuperscript{\emph{c}}     & 0.00  & -2.78   & -3.44 & -0.58  & 0.94  \\ \bottomrule
\end{tabular}%
}     
\small
\textsuperscript{\emph{a}} Within parentheses, the SO corrections (at the CASSCF/aug-cc-pVTZ level) are given. \\  
\textsuperscript{\emph{b}} Relative ZPE corrections at the revDSD-PBEP86-D3(BJ)/jun-cc-pV(T+$d$)Z level. 
\\
\textsuperscript{\emph{c}} Standard state: 1 atm, 298 K; at the revDSD-PBEP86-D3(BJ)/jun-cc-pV(T+$d$)Z level.   
\end{table}

\subsection{The \ce{H2S + Cl} reaction: Rate constants}

The rate constant for the H-abstraction from \ce{H2S} by Cl has been computed at different levels of theories, as described in detail in Section \ref{sec:kinmod}, solving the multi-well one-dimensional master equation using the chemically significant eigenvalues (CSEs) method within the Rice-Ramsperger-Kassel-Marcus (RRKM) approximation, as detailed by Miller and Klippenstein in ref. \citenum{miller2006master}.

Though the use of a master equation model is generally not necessary to evaluate the rate constant of an abstraction reaction, there are several reasons that warrant its employment, as described by Cavallotti \emph{et al}. in ref. \citenum{cavallotti2018estoktp}. In fact, it allows one to properly limit the contributions to the reactive flux from energy states below the asymptotic energies of the fragment, as well as to account for limitations to the reaction fluxes determined by an eventually slow rate of formation of the precursor complex. This is indeed the situation for the present system, for which the flux through the outer transition state, leading to the formation of the precursor complex, and that through the inner transition state, leading to H abstraction and the formation of the product complex, have comparable values. The consequence is that part of the flux passing through the outer transition state is reflected from the inner transition state.

\begin{figure}[ht]
    \centering
    \includegraphics[scale=1.5]{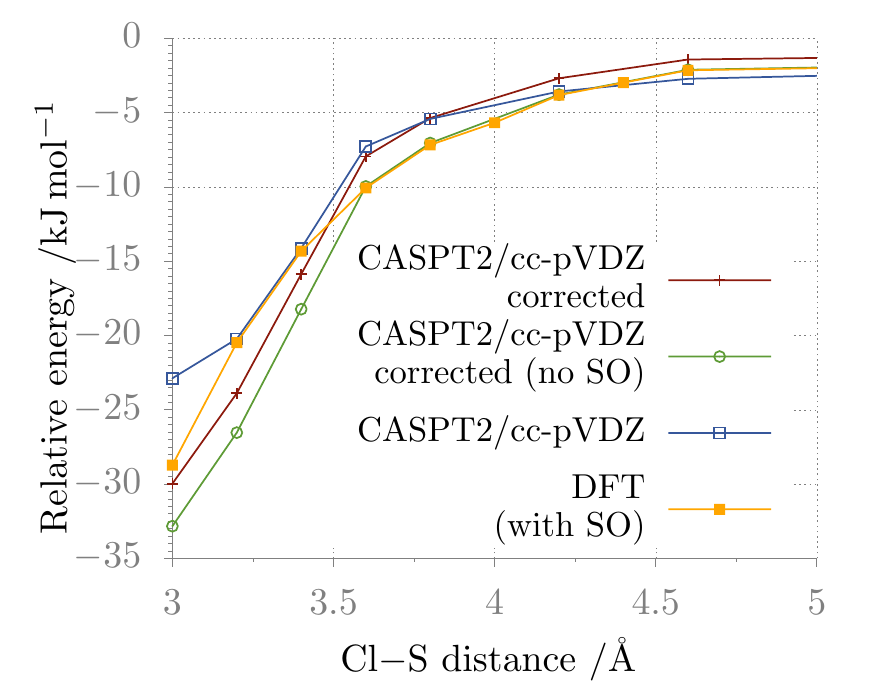}
\caption{Interaction potential between Cl and \ce{H2S} calculated at different levels of theory: the uncorrected CASPT2/cc-pVDZ level; the CASPT2/cc-pVDZ level corrected for geometry relaxation, high level energy contributions, and SO effects (see equation \ref{eq:vrctst}); the CASPT2/cc-pVDZ level corrected for geometry relaxation and high level energy contributions; the revDSD-PBEP86-D3(BJ)/jun-cc-pV(T+$d$)Z corrected for SO effects.}
    \label{fig:entrance}
\end{figure}

As mentioned in Section \ref{sec:kinmod}, reactive fluxes through the outer transition state, which occur on a barrierless MEP, have been computed at three different levels of theory: VRC-TST, VTST, and PST. The \ce{Cl\bond{-}H2S} interaction potential used to determine the VRC-TST flux uncorrected for geometry relaxation, active-space size and basis-set size, and at different levels of corrections is reported in Figure \ref{fig:entrance}. It is interesting to observe that the corrected CASPT2 potential used in VRC-TST calculations is similar, though not exactly overlapped with that determined at the revDSD-PBEP86-D3(BJ)/jun-cc-pV(T+$d$)Z level of theory. This suggests that this functional may be used to perform VRC-TST calculations for open-shell systems for which it may be cumbersome, e.g. for difficulties in converging to a proper active space, to determine the interaction potential at the CASPT2 level. 

\begin{figure}[ht]
    \centering
    \includegraphics[scale=1.5]{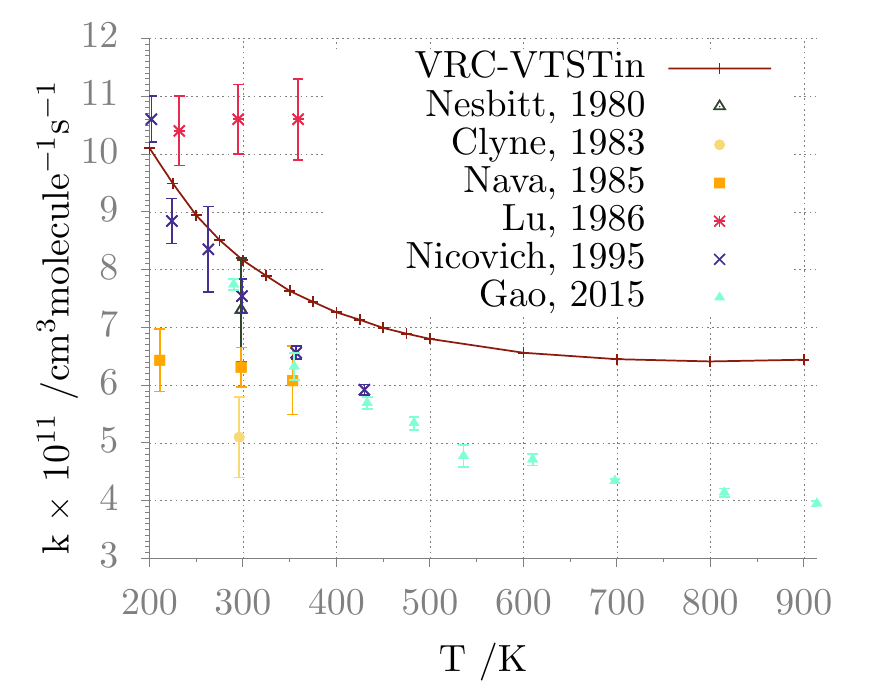}
\caption{The \ce{H2S + Cl} global rate constant: comparison between computed (VRC-TST theory for the outer TS, VTST with vibrational frequencies evaluated using curvilinear internal coordinates for the inner TS, small curvature theory for tunneling) and experimental data.}
    \label{fig:rates}
\end{figure}

As already mentioned, master equation simulations have been performed using the MESS software. The collisional energy transfer probability has been described by means of the single exponential down model\cite{tardy1966collisional} with a temperature dependence $\langle\Delta E\rangle_{\mathrm{down}}$ of 260$(T/298)^{0.875 }$~\SI{}{\per\centi\metre} in an argon bath gas. Different models have been employed to compute reaction fluxes through the inner and the outer transition states. The highest level simulations have been obtained using VRC-TST for the outer TS and VTST in curvilinear internal coordinates for the inner TS (referred as VTSTin), the results being reported in Table \ref{tab:rate} and compared with experimental data in Figure \ref{fig:rates}.

\begin{table}
\caption{Rate coefficients for the \ce{H2S + Cl} reaction at various temperatures (pressure = 1 atm).\textsuperscript{\emph{a}} Values in \SI{}{\cubic\centi\metre\per\molecule\per\second}.}
\label{tab:rate}
\resizebox{0.5\textwidth}{!}{%
\begin{tabular}{@{}cccc@{}}
\toprule
T {[}K{]} & VRC-VTSTin        & VTST-VTSTin           & PST-VTSTin           \\ \midrule
200     & \num{1.01d-10}	&        \num{1.02d-10} & \num{1.31d-10} \\
225     & \num{9.49d-11} 	& 	 \num{9.53d-11} & \num{1.22d-10} \\
250     & \num{8.94d-11} 	& 	 \num{9.03d-11} & \num{1.14d-10} \\
275     & \num{8.51d-11} 	& 	 \num{8.64d-11} & \num{1.08d-10} \\
300     & \num{8.16d-11} 	& 	 \num{8.33d-11} & \num{1.03d-10} \\
325     & \num{7.89d-11} 	& 	 \num{8.07d-11} & \num{9.84d-11} \\
350     & \num{7.63d-11} 	& 	 \num{7.87d-11} & \num{9.50d-11} \\
375     & \num{7.44d-11} 	& 	 \num{7.69d-11} & \num{9.21d-11} \\
400     & \num{7.26d-11} 	& 	 \num{7.56d-11} & \num{8.98d-11} \\
425     & \num{7.13d-11} 	& 	 \num{7.45d-11} & \num{8.78d-11} \\
450     & \num{6.99d-11} 	& 	 \num{7.36d-11} & \num{8.62d-11} \\
475     & \num{6.89d-11} 	& 	 \num{7.29d-11} & \num{8.49d-11} \\ 
500	& \num{6.80d-11}	 &	 \num{7.24d-11}	& \num{8.38d-11} \\
600	& \num{6.56d-11}	 &	 \num{7.17d-11}	& \num{8.16d-11} \\
700	& \num{6.45d-11}	 &	 \num{7.24d-11}	& \num{8.16d-11} \\
800	& \num{6.41d-11}	 &	 \num{7.41d-11}	& \num{8.31d-11} \\
900	& \num{6.44d-11}	 &	 \num{7.65d-11}	& \num{8.57d-11} \\ \bottomrule
\end{tabular}%
}
\small

\textsuperscript{\emph{a}} The various prefixes stand for the theoretical methods used to handle the barrierless entrance channel, while the VTSTin suffix means that the inner TS is handled with VTST in curvilinear internal coordinates. The barrierless exit channel is always treated with PST. \\
\end{table}

\begin{figure}[ht]
    \centering
    \includegraphics[scale=1.5]{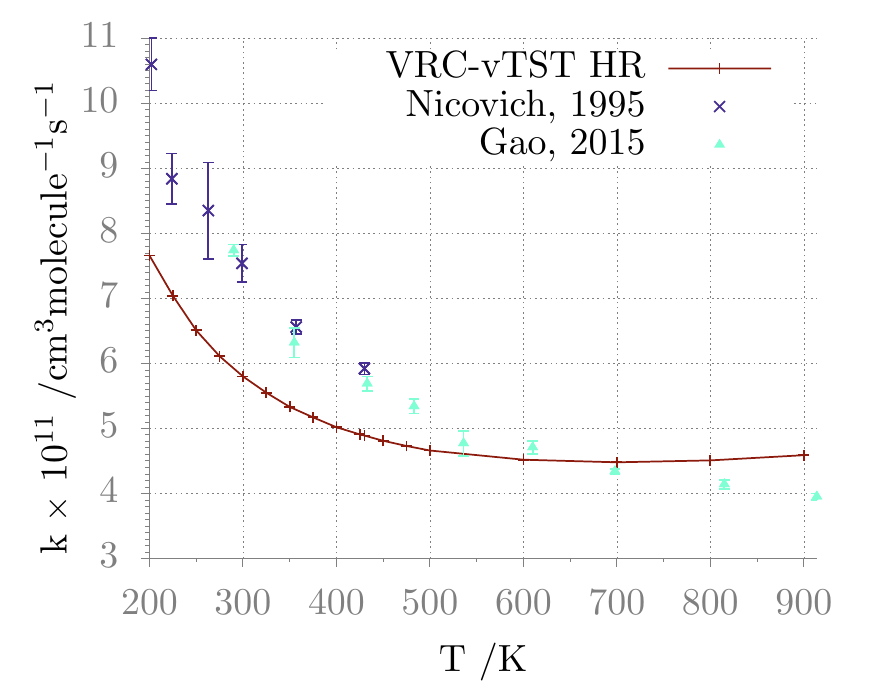}
\caption{The \ce{H2S + Cl} global rate constant: comparison between computed (same level as in Figure \ref{fig:rates}, but modeling of the internal motion for the interconversion between the optical isomers of the TS with a 1D hindered rotor) and experimental data.}
    \label{fig:rate_hr}
\end{figure}

The comparison between calculated and experimental data shows an excellent agreement at 300 K, the temperature at which most measurements were made. Indeed, the calculated \SI{7.76d-11}{\cubic\centi\metre\per\molecule\per\second} value is in quite good agreement with the \SI{7.4d-11}{\cubic\centi\metre\per\molecule\per\second} datum recommended by Atkinson \emph{et al.}\cite{atkinson2004evaluated} on the basis of an extensive review. Furthermore, the calculated rate is in excellent agreement with the rate constant measured in the 200-433 K temperature range by Nicovich \emph{et al.}\cite{nicovich1995kinetics}, from which it differs by about 12\% at most. The calculated global rate constant is almost pressure independent in the considered conditions, which agrees with the experimental data that show that there is no measurable collisional stabilization of the entrance well.\cite{gao2015high} However, the temperature trend is not perfectly reproduced, as the experimental data are slightly underestimated at low temperatures and slightly overestimated at high temperatures. The discrepancy is more evident for the recent measurements by Gao \emph{et al.}\cite{gao2015high}, which are overestimated by a factor of 1.5 at 900 K. Even if such a disagreement is relatively small and it has been observed only with respect to a single set of experimental data, it is anyway useful to try to understand its origin. It is first of all noted that the structure of the saddle point of the inner TS has an optical isomer, which thus effectively doubles the density of states (DOS) of the TS and, consequently, the rate constant. The two optical isomers are separated by a second order saddle point with a barrier of about \SI{10}{\kilo\joule\per\mol}. Therefore, it is likely that, as the temperature increases, the two isomers interconvert among themselves. If this is the case, then the DOS of the TS is overestimated by up to a factor of 2. To investigate whether this can be the case, the 1D PES for the conversion between the two isomers has been determined as a function of the Cl-H-S-H dihedral angle and the partition function of the corresponding vibrational internal motion has been replaced with a 1D hindered rotor model. The rate constants calculated at different temperatures are compared with the experimental temperature-dependent data in Figure \ref{fig:rate_hr}. The computed results are now in quantitative agreement with experiments at high temperature, and differ at most by a factor of 1.38 at 200 K. This outcome confirms, as it is well known in the literature, that one of the key aspects in the estimation of an accurate rate constant using TST is the proper description of anharmonic internal motions, and that it is sometimes necessary to use different models depending on the investigated temperature and pressure conditions.

\begin{figure}[ht]
    \centering
    \includegraphics[scale=1.5]{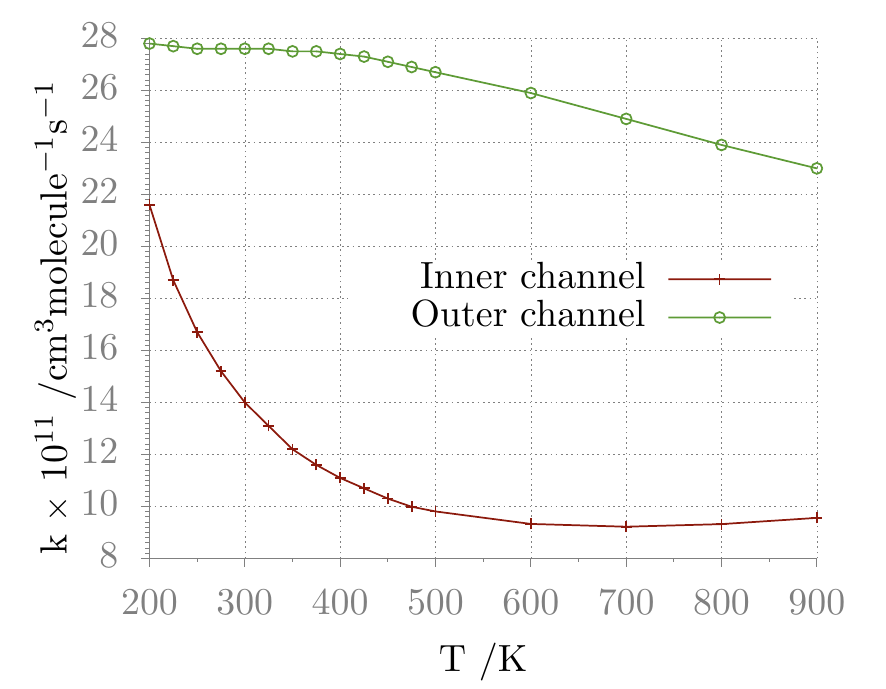}
\caption{Rate constants of the outer and inner channels computed using VRC-TST and VTST theories, respectively.}
    \label{fig:rate_inout}
\end{figure}

In order to compare the contribution of the entrance and inner channels to the global rate constant, it is interesting to report the rates of each channel computed solving the ME fictitiously enhancing the rate of the other channel (Figure \ref{fig:rate_inout}). As it can be observed, both rates exhibit a negative activation energy, in agreement with experimental observations, and their values are comparable, though the rate of the inner channel is smaller, and it thus impacts more significantly the global reaction flux.

\begin{figure}[ht]
    \centering
    \includegraphics[scale=1.5]{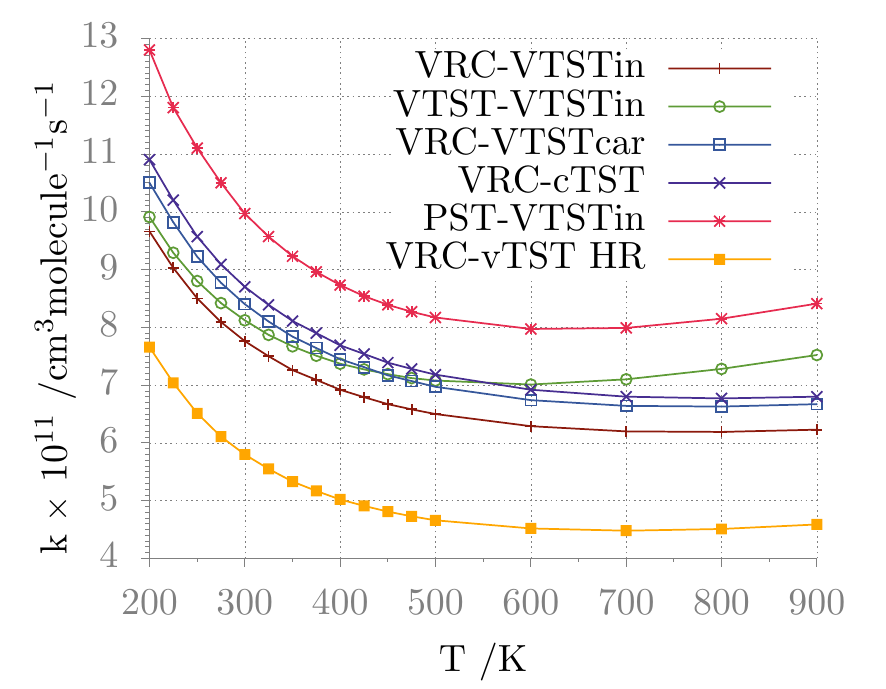}
\caption{Global rate constants computed using different theoretical approaches to determine the fluxes through the inner and outer transition states. The nomenclature is `outer TS - inner TS’: 1) the outer TS flux: VRC-TST (VRC), VTST (VTST), or PST (PST); 2) the inner TS flux: the VTST level with vibrational frequencies computed using internal curvilinear coordinates (VTSTin) or Cartesian coordinates (VTSTcar), conventional TST (cTST), VTST with one internal mode modelled as a 1D hindered rotor (vTST HR).}
    \label{fig:rate_theo}
\end{figure}

The impact of the level of theory chosen to compute the inner and outer TS fluxes on the global rate constant is analyzed in Figure \ref{fig:rate_theo}.
It can be observed that, for the outer channel, VRC-TST and VTST give similar results that differ at most by a factor of 1.1, while PST predictions deviate by up to a factor of 1.2. Despite this, it is interesting to notice that there is a slight but significant qualitative difference in the temperature dependence between the rate constants computed using VRC-TST, in better agreement with the experimental trend, and those determined at the other theoretical levels. It should also be recalled that the global rate constant is mostly controlled by the rate of the inner TS, so that the differences between the levels of theory used for the outer TSs are mitigated. The analysis of the impact of the chosen theoretical level for the inner TS shows that variational effects have a minor impact, though the rate constant computed using the internal coordinate model is in better agreement with experimental data. The most relevant effect on the rate constant, as commented above, is given by the use of the 1D hindered rotor model. Finally, though not shown, it has been found that using the Eckart model rather than small curvature theory to compute the tunnelling contributions has a negligible impact on the rate constant evaluation. The reason is that tunnelling corrections are small for this system, because the RW adduct is not significantly collisionally stabilized in the examined temperature and pressure conditions and the energy barrier is submerged with respect to reactants.

It can be concluded that the elementary processes that contribute to the reactive fluxes change depending on the ranges of temperatures and pressures that are investigated, and that their rate must be determined at a suitable level of theory in order to obtain quantitative agreement with experimental data. For example, canonical VTST is not apt to study the entrance channel for this system, not much because it assumes a thermal distribution (thus allowing for computing a canonical $k(T)$ rate constant instead of the microcanonical E, or E,J resolved rates, $k(E)$ or $k(E,J)$), but rather because, as it is often implemented in the literature, it uses the harmonic approximation to evaluate reactive fluxes along the minimum energy path, which is improper for loosely interacting fragments. A more `proper' evaluation of the reactive fluxes is that given by VRC-TST.

\section{Conclusions}

The kinetics of radical-molecule reactions is of remarkable interest in several fields including, inter alia, atmospheric- and astro-chemistry. However, obtaining quantitative rate constants for such reactions by means of theoretical methods is challenging because of the difficulties that can be faced in the accurate description of some stationary points (intermediates and/or transition states). Indeed, they might show strong correlation effects, the situation being more involved when third-row atoms are present. Furthermore, SO coupling might be relevant for open-shell species. On these grounds, the first aim of this paper was to investigate a prototypical reaction of this kind, namely the \ce{H2S + Cl} addition/elimination reaction, beyond the usual ``gold standard'' of quantum-chemical calculations, represented by CCSD(T) possibly including the extrapolation to the CBS limit. To this aim, a HEAT-like approach, which includes the full treatment of triple and quadruple excitations together with diagonal Born-Oppenheimer corrections and relativistic effects, combined with a proper treatment of the SO coupling has been employed. This level of theory, in conjunction with anharmonic ZPE corrections evaluated using the double-hybrid revDSD-PBEP86-D3(BJ) functional in the framework of the VPT2 model, is expected to fulfil a sub-\SI{}{\kilo\joule\per\mol} accuracy, thus allowing an unbiased analysis of the ability of different kinetic models in reproducing the experimental reaction rates. In the present work, different approaches of increasing accuracy have been employed to describe the barrierless entrance channel of the reaction, whose role (in evaluating the global reaction rate) depends on the examined temperature and pressure conditions.

In this connection, even the quite simple PST leads to results within a factor of two with respect to their experimental counterparts, whereas the more refined VTST and, especially, VRC-TST models lead to results in quantitative agreement with experiment. These outcomes show unambiguously that this reaction can be well described by models based on the transition state theory, provided that the underlying electronic structure computations are sufficiently accurate and that barrierless channels are properly described. 

Furthermore, in view of extending the accuracy reached by our approach to reactive PESs involving larger systems, we have tested the performance of computationally less expensive composite schemes, which would become indeed unavoidable in such cases. Our conclusion is that different variants of the so-called ``cheap'' approach perform remarkably well. At the same time, last-generation double-hybrid functionals can be profitably used to optimize geometries and evaluate vibrational contributions. In summary, in our opinion, a promising route for computing reaction rates in semi-quantitative agreement with experiment (i.e. well within a factor of two) for quite large molecular systems can be based on ChS energy evaluations of the relevant stationary points coupled to TST for the activated steps and to PST for barrierless steps. 
All these ingredients must be finally introduced in a master equation model of the overall reaction network, which must include all the elementary processes that can contribute to the reactive fluxes, with rates computed at a suitable level of theory.
\begin{acknowledgement}
This work has been supported by MIUR 
(Grant Number 2017A4XRCA) and by the University of Bologna (RFO funds). The SMART@SNS Laboratory (http://smart.sns.it) is acknowledged for providing high-performance computing facilities. JL thanks Francesca Montanaro for the help in the elaboration of the graphical abstract.
\end{acknowledgement}

\begin{suppinfo}
\begin{itemize}
\item Input file (example) for the RRKM-ME calculation of the rate constant for the \ce{H2S + Cl} reaction using the MESS package.
\item Thermochemistry results
\end{itemize}
\end{suppinfo}

\bibliography{achemso-demo}

\end{document}


\clearpage

\section*{Thermochemistry}

Using the electronic structure calculations carried out and literature data, the heats of formation (at 298 K, 1 atm) of \ce{H2S}, HS, and HCl have been derived. This only required additional computations for the sulfur atom ($^3P$).

\begin{table}[ht!]                                      
\caption{Thermochemistry.\textsuperscript{\emph{a}}}       
\label{tab:thermo}                                           
\resizebox{\textwidth}{!}{%
\begin{tabular}{@{}lccccccc@{}}                                         
\toprule                                                                  
        &$E_{el}$\textsuperscript{\emph{b}}&$\Delta E_{el}$\textsuperscript{\emph{b}}& ZPE\textsuperscript{\emph{c}}& $H^\circ - H_\circ^\circ$  & SO\textsuperscript{\emph{d}}& $\Delta_f H^\circ$ (calc)  &$\Delta_f H^\circ$ (exp)         \\ \midrule     
        &   (Hartree)                      & (\SI{}{\kilo\joule\per\mol})            & (\SI{}{\kilo\joule\per\mol}) &(\SI{}{\kilo\joule\per\mol})& (\SI{}{\kilo\joule\per\mol})&(\SI{}{\kilo\joule\per\mol})&(\SI{}{\kilo\joule\per\mol})     \\ \cmidrule(l){2-8}            
H       &          -0.5                    &    0.0                                  &      0.0                     &  6.20\textsuperscript{\emph{e}}                      &  0.0                        &  --                        & 218.00\textsuperscript{\emph{e}} \\                                                                                                                                     
S       &        -399.1715515              &    0.0                                  &      0.0                     &  6.66\textsuperscript{\emph{e}}                      & -2.16                       &  --                        & 277.17\textsuperscript{\emph{e}} \\                                                                                                                                     
Cl      &        -461.5429218              &    0.0                                  &      0.0                     &  6.20                      & -3.28                       &  --                        & 121.30\textsuperscript{\emph{f}} \\                                                                                                                                     
HS      &        -399.8107595              & -365.46                                 &     16.14                    &  8.81                      & -2.10                       & 141.86                     & 141.87\textsuperscript{\emph{e}} \\                                                                                                                                     
HCl     &        -462.2138181              & -448.66                                 &     17.88                    &  8.78                      &  0.0                        & -91.82                     & -92.31\textsuperscript{\emph{f}} \\                                                                                                                                     
\ce{H2S}&        -400.4635863              & -766.68                                 &     39.63                    & 10.44                      &  0.0                        & -20.34                     & -20.50\textsuperscript{\emph{f}} \\ \bottomrule                                                           
\end{tabular}
}     
\textsuperscript{\emph{a}} Standard state: 1 atm, 298 K.   
\textsuperscript{\emph{b}} At the CBS+CV+DBOC+rel+fT+fQ level. \\
\textsuperscript{\emph{c}} At the the revDSD-PBEP86-D3(BJ)/jun-cc-pV(T+$d$)Z level. \\
\textsuperscript{\emph{d}} At the CASSCF/aug-cc-pVTZ level. 
\textsuperscript{\emph{e}} Ref.~\citenum{hsnew}. 
\textsuperscript{\emph{f}} Ref.~\citenum{codata}.   
\end{table}

\bibliography{achemso-demo}

\clearpage

\section*{Input file for RRKM-ME calculations using the MESS package for the \ce{H2S + Cl} reaction}

In the following, the input file for the VRC-VTSTin calculation is reported. All the important molecular properties are enclosed. The nomenclature used is the same used in Section 3.4.

\subsection*{VRC-VTSTin}
\begin{tiny}
\begin{verbatim}
!***************************************************
!               GLOBAL SECTION
!***************************************************
!!!!!!!!!!!!!!!!!!!!!!!!!!!!!!!!!!!!!!!!!!!!!!!!!!!!
!
!
TemperatureList[K]                      200	225	250	275	300	325	350	375	400   425 430 450 475 500 600 700 800 900 1000	
PressureList[atm]                       0.5 1 
!
!
EnergyStepOverTemperature               .2         ! [Discretization energy step (global relax matrix)] / T                   
ExcessEnergyOverTemperature             30         ! [Highest barrier in the model (global relax matrix)] / T                      
ModelEnergyLimit[kcal/mol]              400        ! Highest reference energy used in the calculation ( or ReferenceEnergy[kcal/mol])
!
CalculationMethod                       direct     ! direct or low-eigenvalue                     
!
WellCutoff                              20         ! well truncation parameter : Max { dissociation limit (min barrier rel. to bottom of the well) / T }
ChemicalEigenvalueMax                   0.2        ! Max chemical eigenvalue / Lowest Collision relaxation eigenvalue 
!
ReductionMethod                         diagonalization ! [low eigenvalue method only] diagonalization or projection (default)
!      
!!!!!!!!!test!!!!!!!!!!!!!!!!!!!!
!WellCutoff                            10
!ChemicalEigenvalueMin                 1.e-6          #only for direct diagonalization method
!!!!!!!!test!!!!!!!!!!!!!!!!!!!!!!!!!!
AtomDistanceMin[bohr]                  1.3
!!
RateOutput                              rate_VRC.out                        ! output file name for rate coefficients                         
!
!
!!!!!!!!!!!!!!!!!!!!!!!!!!!!!!!!!!!!!!!!!!!!!!!!!!!!
!***************************************************
!               MODEL SECTION
!***************************************************
!!!!!!!!!!!!!!!!!!!!!!!!!!!!!!!!!!!!!!!!!!!!!!!!!!!!
!
!                          
Model        
!                                                           
  EnergyRelaxation                                                      ! Default collisional energy relaxation kernel                              
    Exponential                                                         ! Currently the only possible energy relaxation model              
       Factor[1/cm]                     260                             ! (Delta_E_down)^(0) @ standard T (300 K)                          
       Power                            0.875                           ! Power n in the expression (Delta_E_down) = (Delta_E_down)^(0) (T/T0)^(n)
       ExponentCutoff                   10                              ! if (Delta_E) / (Delta_E_down) > value  transition probability is zero     
    End  
!                                                               
  CollisionFrequency                                                    ! Collision frequency model
    LennardJones                                                        ! Currently the only possible collisional frequency model  based on LJ potential
       Epsilons[K]                       90.58  617.0                   ! Epsilon_1 and Epsilon_2      
       Sigmas[angstrom]                  3.54    5.62                   ! Sigma_1 and Sigma_2 
       Masses[amu]                       39.948 69.0                    ! Masses of the buffer gas molecule and of the complex (check order) 
   End      
!
!*************************************************
!
!***************************************************
!!!!!!!!!!!!!!!!!!!!!!!!!!!!!!!!!!!!!!!!!!!!!!!!!!!!
!***************************************************
!  REACTANTS					
!***************************************************


 Bimolecular REACS
 Fragment REACT1
 RRHO     
 Geometry[angstrom]               3
   S                  0.00000000    0.00000000    0.10283700
   H                  0.00000000    0.96718500   -0.82269200
   H                  0.00000000   -0.96718500   -0.82269200
 	Core 	RigidRotor
 		SymmetryFactor    2.0000000000000000     
 	End
     Frequencies[1/cm]            3
   1217.89 2735.01 2750.20
  ZeroEnergy[kcal/mol]             0.
  ElectronicLevels[1/cm]                      1
   0            1
 !************************************
 End
!*********
 Fragment REACT2
  Atom
  Name Cl
  ElectronicLevels[1/cm]                      2
   0       4
   881.00  2     
  
 !************************************
 End
!*********
GroundEnergy[kcal/mol] 0.0
End
 Bimolecular PRODS
 Fragment PROD1
 RRHO     
 Geometry[angstrom]               2
   H                  0.00000000    0.00000000   -1.26323400
   S                  0.00000000    0.00000000    0.07895200
 	Core 	RigidRotor
 		SymmetryFactor    1.0000000000000000     
 	End
     Frequencies[1/cm]            1
   2719.96 
  ZeroEnergy[kcal/mol]             0.
  ElectronicLevels[1/cm]                      2
   0.0000000000000000        2.0000000000000000     
   376.95999999999998        2.0000000000000000     
 !************************************
 End
!*********
 Fragment PROD2
 RRHO     
 Geometry[angstrom]               2
   H                  0.00000000    0.00000000   -1.20561900
   Cl                 0.00000000    0.00000000    0.07091900
 	Core 	RigidRotor
 		SymmetryFactor    1.0000000000000000     
 	End
     Frequencies[1/cm]            1
   3006.51
  ZeroEnergy[kcal/mol]             0.
  ElectronicLevels[1/cm]                      1
   0.0000000000000000        1.0000000000000000     
  
 !************************************
 End
GroundEnergy[kcal/mol]   -12.39
End
!*********
 Well WR
 Species
 RRHO     ! well
 Geometry[angstrom]               4
   H                  0.86853200   -1.21794200    0.97069800
   S                 -0.05263800   -1.26295000    0.00000000
   H                  0.86853200   -1.21794200   -0.97069800
   Cl                -0.05263800    1.33194700    0.00000000
 	Core 	RigidRotor
 		SymmetryFactor    1.0000000000000000     
 	End
     Frequencies[1/cm]            6
   235.50 361.30 394.60 1211.08 2735.16 2749.92
  ZeroEnergy[kcal/mol]     -8.68 
  ElectronicLevels[1/cm]                      1
   0.0000000000000000        2.0000000000000000     
 End 
 End 
 !************************************
 Well WP
 Species
 RRHO     ! well
 Geometry[angstrom]               4
    S                 -1.89837200   -0.07864900   -0.00000300
    H                 -1.96120000    1.26293700   -0.00000900
    H                  0.58255000   -0.07505800    0.00001500
    Cl                 1.86780000    0.00414800    0.00000200
  	Core 	RigidRotor
  		SymmetryFactor    1.0000000000000000     
 	End
    Frequencies[1/cm]            6
   92.55 144.45 254.96 397.14 2715.09 2844.56
  ZeroEnergy[kcal/mol]              -13.91
  ElectronicLevels[1/cm]                      1
   0.0000000000000000        2.0000000000000000     
 End 
 End 
 !************************************
 Barrier	B1  REACS WR
 RRHO
    Stoichiometry H2S1Cl1                       
     Core Rotd
       File flux_lr_so9.dat
       SymmetryFactor    2.2200000000     
     End
     Frequencies[1/cm]            3
   1217.89 2735.01 2750.20
    ZeroEnergy[kcal/mol]    0.0
     ElectronicLevels[1/cm]               1
   0.0000000000000000        2.0000000000000000     
 End
 !************************************
 Barrier	B3  WP PRODS
   RRHO 
    Stoichiometry  H2S1Cl1                       
    Core PhaseSpaceTheory
      FragmentGeometry[angstrom]               2
   H                  0.00000000    0.00000000   -1.26323400
   S                  0.00000000    0.00000000    0.07895200
      FragmentGeometry[angstrom]               2
   H                  0.00000000    0.00000000   -1.20561900
   Cl                 0.00000000    0.00000000    0.07091900
 		SymmetryFactor     1.0000000000000000     
          PotentialPrefactor[au] 	230.
          PotentialPowerExponent 	6
 	    End
    Frequencies[1/cm]            2
    2719.96 3006.51 
ZeroEnergy[kcal/mol]    -12.39        
 ElectronicLevels[1/cm]                   2
  0.0000000000000000        2.0000000000000000     
  376.95999999999998        2.0000000000000000     
End
 !************************************
 Barrier B2 WR WP
 Variational
 RRHO            !            90
 Geometry[angstrom]            4
 Cl   1.36787   0.01974  -0.00798                                               
 H   -0.27379  -0.19797   0.71945                                               
 S   -1.44332  -0.05571  -0.01648                                               
 H   -1.39984   1.28048   0.08033                                               
 	Core 	RigidRotor
           SymmetryFactor   0.50000000000000000     
 End
     Frequencies[1/cm]            5
   2733.8      2093.0      1136.8      394.25      189.40    
 ZeroEnergy[kcal/mol]   -1.53458070657077     
  ElectronicLevels[1/cm]                      1
   0.0000000000000000        2.0000000000000000     
 End
 !*********************************************
 RRHO            !            91
 Geometry[angstrom]            4
 Cl   1.36762   0.01969  -0.00777                                               
 H   -0.26174  -0.19606   0.70931                                               
 S   -1.44342  -0.05572  -0.01638                                               
 H   -1.39997   1.28056   0.07988                                               
 	Core 	RigidRotor
           SymmetryFactor   0.500000000000000000     
 End
     Frequencies[1/cm]            5
   2733.4      2039.3      1129.8      396.99      191.83    
 ZeroEnergy[kcal/mol]   -1.46672132293552     
  ElectronicLevels[1/cm]                      1
   0.0000000000000000        2.0000000000000000     
 End
 !*********************************************
 RRHO            !            92
 Geometry[angstrom]            4
 Cl   1.36736   0.01964  -0.00756                                               
 H   -0.24947  -0.19415   0.69944                                               
 S   -1.44353  -0.05573  -0.01629                                               
 H   -1.40008   1.28065   0.07946                                               
 	Core 	RigidRotor
           SymmetryFactor   0.50000000000000000     
 End
     Frequencies[1/cm]            5
   2733.0      1982.1      1122.1      399.78      194.59    
 ZeroEnergy[kcal/mol]   -1.40962588916224     
  ElectronicLevels[1/cm]                      1
   0.0000000000000000        2.0000000000000000     
 End
 !*********************************************
 RRHO            !            93
 Geometry[angstrom]            4
 Cl   1.36710   0.01959  -0.00735                                               
 H   -0.23696  -0.19225   0.68988                                               
 S   -1.44364  -0.05574  -0.01620                                               
 H   -1.40016   1.28073   0.07908                                               
 	Core 	RigidRotor
           SymmetryFactor   0.50000000000000000     
 End
     Frequencies[1/cm]            5
   2732.5      1921.6      1113.6      402.59      197.68    
 ZeroEnergy[kcal/mol]   -1.36299419518592     
  ElectronicLevels[1/cm]                      1
   0.0000000000000000        2.0000000000000000     
 End
 !*********************************************
 RRHO            !            94
 Geometry[angstrom]            4
 Cl   1.36684   0.01954  -0.00714                                               
 H   -0.22418  -0.19036   0.68068                                               
 S   -1.44375  -0.05574  -0.01612                                               
 H   -1.40021   1.28081   0.07873                                               
 	Core 	RigidRotor
           SymmetryFactor   0.500000000000000000     
 End
     Frequencies[1/cm]            5
   2732.1      1858.4      1104.3      405.40      201.13    
 ZeroEnergy[kcal/mol]   -1.3259970893985     
  ElectronicLevels[1/cm]                      1
   0.0000000000000000        2.0000000000000000     
 End
 !*********************************************
 RRHO            !            95
 Geometry[angstrom]            4
 Cl   1.36658   0.01949  -0.00694                                               
 H   -0.21112  -0.18850   0.67187                                               
 S   -1.44388  -0.05575  -0.01606                                               
 H   -1.40024   1.28089   0.07841                                               
 	Core 	RigidRotor
           SymmetryFactor   0.50000000000000000     
 End
     Frequencies[1/cm]            5
   2731.6      1793.2      1094.0      408.20      204.92    
 ZeroEnergy[kcal/mol]   -1.30378030473142     
  ElectronicLevels[1/cm]                      1
   0.0000000000000000        2.0000000000000000     
 End
 !*********************************************
 RRHO            !            96
 Geometry[angstrom]            4
 Cl   1.36633   0.01944  -0.00674                                               
 H   -0.19777  -0.18666   0.66349                                               
 S   -1.44402  -0.05576  -0.01600                                               
 H   -1.40024   1.28097   0.07812                                               
 	Core 	RigidRotor
           SymmetryFactor   0.50000000000000000     
 End
     Frequencies[1/cm]            5
   2731.1      1727.3      1082.9      410.98      209.05    
 ZeroEnergy[kcal/mol]   -1.28852480910309     
  ElectronicLevels[1/cm]                      1
   0.0000000000000000        2.0000000000000000     
 End
 !*********************************************
 RRHO            !            97
 Geometry[angstrom]            4
 Cl   1.36608   0.01939  -0.00655                                               
 H   -0.18414  -0.18488   0.65558                                               
 S   -1.44418  -0.05577  -0.01596                                               
 H   -1.40022   1.28105   0.07786                                               
 	Core 	RigidRotor
           SymmetryFactor   0.50000000000000000     
 End
     Frequencies[1/cm]            5
   2730.6      1661.8      1070.8      413.69      213.48    
 ZeroEnergy[kcal/mol]   -1.29115107994113     
  ElectronicLevels[1/cm]                      1
   0.0000000000000000        2.0000000000000000     
 End
 !*********************************************
 RRHO            !            98
 Geometry[angstrom]            4
 Cl   1.36583   0.01935  -0.00636                                               
 H   -0.17022  -0.18315   0.64815                                               
 S   -1.44435  -0.05577  -0.01593                                               
 H   -1.40017   1.28112   0.07763                                               
 	Core 	RigidRotor
           SymmetryFactor   0.50000000000000000     
 End
     Frequencies[1/cm]            5
   2730.1      1598.0      1057.9      416.27      218.14    
 ZeroEnergy[kcal/mol]   -1.30368283227277     
  ElectronicLevels[1/cm]                      1
   0.0000000000000000        2.0000000000000000     
 End
 !*********************************************
 RRHO            !            99
 Geometry[angstrom]            4
 Cl   1.36560   0.01930  -0.00617                                               
 H   -0.15606  -0.18147   0.64119                                               
 S   -1.44454  -0.05578  -0.01591                                               
 H   -1.40011   1.28119   0.07742                                               
 	Core 	RigidRotor
           SymmetryFactor   0.50000000000000000     
 End
     Frequencies[1/cm]            5
   2729.6      1537.3      1044.1      418.64      222.94    
 ZeroEnergy[kcal/mol]   -1.33721209213419     
  ElectronicLevels[1/cm]                      1
   0.0000000000000000        2.0000000000000000     
 End
 !*********************************************
 RRHO            !           100
 Geometry[angstrom]            4
 Cl   1.36537   0.01926  -0.00599                                               
 H   -0.14168  -0.17985   0.63466                                               
 S   -1.44475  -0.05578  -0.01589                                               
 H   -1.40002   1.28126   0.07723                                               
 	Core 	RigidRotor
           SymmetryFactor   0.5000000000000000     
 End
     Frequencies[1/cm]            5
   2729.1      1482.5      1032.3      423.97      228.94    
 ZeroEnergy[kcal/mol]   -1.37987486475066     
  ElectronicLevels[1/cm]                      1
   0.0000000000000000        2.0000000000000000     
 End
 !*********************************************
 RRHO            !           101
 Geometry[angstrom]            4
 S    0.00000   0.00000   0.00000                                               
 H    0.00000   0.00000   1.34174                                               
 H    1.48190   0.00000  -0.02582                                               
 Cl   2.55774  -1.20674   0.18741                                               
 	Core 	RigidRotor
           SymmetryFactor   0.50000000000000000     
 End
     Frequencies[1/cm]            5
   2728.7      1427.7      1014.0      422.35      232.63    
 ZeroEnergy[kcal/mol]   -1.43880326244397     
  ElectronicLevels[1/cm]                      1
   0.0000000000000000        2.0000000000000000     
 End
 !*********************************************
 RRHO            !           102
 Geometry[angstrom]            4
 Cl   1.36495   0.01917  -0.00564                                               
 H   -0.11227  -0.17684   0.62311                                               
 S   -1.44523  -0.05579  -0.01590                                               
 H   -1.39980   1.28137   0.07693                                               
 	Core 	RigidRotor
           SymmetryFactor   0.50000000000000000     
 End
     Frequencies[1/cm]            5
   2728.3      1375.8      995.55      420.71      235.90    
 ZeroEnergy[kcal/mol]   -1.51855761905849     
  ElectronicLevels[1/cm]                      1
   0.0000000000000000        2.0000000000000000     
 End
 !*********************************************
 RRHO            !           103
 Geometry[angstrom]            4
 Cl   1.36477   0.01913  -0.00547                                               
 H   -0.09731  -0.17545   0.61803                                               
 S   -1.44550  -0.05579  -0.01592                                               
 H   -1.39967   1.28141   0.07682                                               
 	Core 	RigidRotor
           SymmetryFactor   0.50000000000000000     
 End
     Frequencies[1/cm]            5
   2727.8      1337.9      981.72      424.41      241.57    
 ZeroEnergy[kcal/mol]   -1.61656470546575     
  ElectronicLevels[1/cm]                      1
   0.0000000000000000        2.0000000000000000     
 End
 !*********************************************
 RRHO            !           104
 Geometry[angstrom]            4
 Cl   1.36459   0.01909  -0.00531                                               
 H   -0.08224  -0.17411   0.61327                                               
 S   -1.44579  -0.05579  -0.01595                                               
 H   -1.39952   1.28145   0.07671                                               
 	Core 	RigidRotor
           SymmetryFactor   0.5000000000000000     
 End
     Frequencies[1/cm]            5
   2727.5      1299.2      964.92      424.74      245.39    
 ZeroEnergy[kcal/mol]   -1.74271643276748     
  ElectronicLevels[1/cm]                      1
   0.0000000000000000        2.0000000000000000     
 End
 !*********************************************
 RRHO            !           105
 Geometry[angstrom]            4
 Cl   1.36443   0.01905  -0.00515                                               
 H   -0.06706  -0.17283   0.60886                                               
 S   -1.44609  -0.05579  -0.01598                                               
 H   -1.39937   1.28148   0.07662                                               
 	Core 	RigidRotor
           SymmetryFactor   0.50000000000000000     
 End
     Frequencies[1/cm]            5
   2727.2      1264.2      947.82      424.63      248.54    
 ZeroEnergy[kcal/mol]   -1.89699850524631     
  ElectronicLevels[1/cm]                      1
   0.0000000000000000        2.0000000000000000     
 End
 !*********************************************
 RRHO            !           106
 Geometry[angstrom]            4
 Cl   1.36428   0.01902  -0.00499                                               
 H   -0.05180  -0.17160   0.60478                                               
 S   -1.44642  -0.05578  -0.01602                                               
 H   -1.39920   1.28150   0.07652                                               
 	Core 	RigidRotor
           SymmetryFactor   0.50000000000000000     
 End
     Frequencies[1/cm]            5
   2726.9      1232.5      930.41      424.04      250.93    
 ZeroEnergy[kcal/mol]   -2.08036873596672     
  ElectronicLevels[1/cm]                      1
   0.0000000000000000        2.0000000000000000     
 End
 !*********************************************
 RRHO            !           107
 Geometry[angstrom]            4
 Cl   1.36415   0.01898  -0.00484                                               
 H   -0.03646  -0.17044   0.60103                                               
 S   -1.44676  -0.05578  -0.01606                                               
 H   -1.39902   1.28152   0.07644                                               
 	Core 	RigidRotor
           SymmetryFactor   0.5000000000000000     
 End
     Frequencies[1/cm]            5
   2726.6      1204.1      912.74      423.02      252.55    
 ZeroEnergy[kcal/mol]   -2.08036873596672     
  ElectronicLevels[1/cm]                      1
   0.0000000000000000        2.0000000000000000     
 End
 !*********************************************
 RRHO            !           108
 Geometry[angstrom]            4
 Cl   1.36403   0.01895  -0.00469                                               
 H   -0.02104  -0.16934   0.59760                                               
 S   -1.44712  -0.05578  -0.01611                                               
 H   -1.39884   1.28153   0.07636                                               
 	Core 	RigidRotor
           SymmetryFactor   0.50000000000000000     
 End
     Frequencies[1/cm]            5
   2726.4      1179.1      894.84      421.60      253.56    
 ZeroEnergy[kcal/mol]   -2.51642042707905     
  ElectronicLevels[1/cm]                      1
   0.0000000000000000        2.000000000000000     
 End
 !*********************************************
 RRHO            !           109
 Geometry[angstrom]            4
 Cl   1.36393   0.01891  -0.00455                                               
 H   -0.00556  -0.16831   0.59449                                               
 S   -1.44751  -0.05578  -0.01617                                               
 H   -1.39864   1.28154   0.07628                                               
 	Core 	RigidRotor
           SymmetryFactor   0.5000000000000000     
 End
     Frequencies[1/cm]            5
   2726.2      1158.1      876.80      419.86      254.22    
 ZeroEnergy[kcal/mol]   -2.77406177607644     
  ElectronicLevels[1/cm]                      1
   0.0000000000000000        2.0000000000000000     
 End
 !*********************************************
 RRHO            !           110
 Geometry[angstrom]            4
 Cl   1.36384   0.01888  -0.00441                                               
 H    0.00998  -0.16735   0.59173                                               
 S   -1.44791  -0.05577  -0.01623                                               
 H   -1.39844   1.28154   0.07621                                               
 	Core 	RigidRotor
           SymmetryFactor   0.5000000000000000     
 End
     Frequencies[1/cm]            5
   2726.0      1142.0      858.69      417.86      254.79    
 ZeroEnergy[kcal/mol]   -3.0462253979558     
  ElectronicLevels[1/cm]                      1
   0.0000000000000000        2.0000000000000000     
 End
 !*********************************************
!  Tunneling    Eckart
!    ImaginaryFrequency[1/cm]    1076.56     
!    WellDepth[kcal/mol]           7.35
!    WellDepth[kcal/mol]           12.65
!    End

       Tunneling    Read
    CutoffEnergy[1/cm]       3500
    ImaginaryFrequency[1/cm]      1076.5599999999999     
    File imactint.dat
    End
 !*********************************************
 End 
  End 





\end{verbatim}
\end{tiny}


\subsubsection*{VRC-TST reactive flux: flux\_lr\_so9.dat}
\begin{tiny}
\begin{verbatim}
              0        7388.88
        13.9008          16396
        28.4966        30980.7
        43.8222        52764.1
        59.9141        83468.9
        76.8106         125276
         94.552         180699
         113.18         252106
         132.74         342081
        153.278         452744
        174.843         580815
        197.486         725672
        221.261         889861
        246.224    1.08211e+06
        272.436    1.30653e+06
        299.959    1.56775e+06
        328.858    1.87055e+06
        359.201    2.21851e+06
        391.062    2.62067e+06
        424.516    3.08577e+06
        459.643    3.62356e+06
        496.526     4.2451e+06
        535.253    4.96392e+06
        575.916    5.79245e+06
        618.613     6.7453e+06
        663.444    7.84349e+06
        710.517    9.08811e+06
        759.944    1.05128e+07
        811.842    1.21388e+07
        866.335     1.3987e+07
        923.552    1.60929e+07
        983.631    1.84841e+07
        1046.71    2.12299e+07
        1112.95    2.43237e+07
         1182.5    2.78659e+07
        1255.52    3.14448e+07
         1332.2    3.48072e+07
        1412.71     3.8266e+07
        1497.25    4.19167e+07
        1586.01    4.56694e+07
        1679.21    4.96416e+07
        1777.07     5.3989e+07
        1879.83    5.86709e+07
        1987.72    6.37684e+07
        2101.01     6.9319e+07
        2219.96    7.53693e+07
        2344.86    8.20702e+07
           2476    8.95202e+07
         2613.7    9.78191e+07
        2758.29    1.07099e+08
         2910.1    1.17503e+08
        3069.51    1.29187e+08
        3236.88    1.42265e+08
        3412.63    1.56895e+08
        3597.16    1.73316e+08
        3790.92    1.91763e+08
        3994.37    2.12575e+08
        4207.99    2.36086e+08
        4432.29    2.62665e+08
         4667.8    2.92709e+08
        4915.09    3.26624e+08
        5174.75     3.6492e+08
        5447.39    4.08341e+08
        5733.66    4.57709e+08
        6034.24    5.13803e+08
        6349.85    5.77476e+08
        6681.25    6.49699e+08
        7029.21    7.31552e+08
        7394.57    8.24541e+08
         7778.2    9.30513e+08
        8181.01    1.05119e+09
        8603.96    1.18833e+09
        9048.06    1.34386e+09
        9514.36    1.52014e+09
          10004    1.72116e+09
        10518.1    1.94994e+09
        11057.9    2.20985e+09
        11624.7    2.50463e+09
        12219.8    2.83886e+09
        12844.7    3.21886e+09
        13500.8    3.65113e+09
        14189.8     4.1418e+09
        14913.2    4.69888e+09
        15672.7    5.32929e+09
        16470.3    6.04446e+09
        17307.7    6.85799e+09
          18187    7.78227e+09
        19110.2    8.82912e+09
        20079.6    1.00109e+10
        21097.5    1.13532e+10
        22166.3    1.28807e+10
        23288.5     1.4614e+10
        24466.8     1.6568e+10
        25704.1    1.87764e+10
        27003.2     2.1293e+10
        28367.2    2.41533e+10
        29799.5    2.73799e+10
        31303.4    3.10112e+10
        32882.5    3.51457e+10
        34540.5    3.98525e+10
        36281.4    4.51663e+10
        38109.4    5.11387e+10
        40028.7    5.79183e+10
        42044.1    6.56466e+10
        44160.2    7.43746e+10
        46382.1    8.41946e+10
        48715.1     9.5299e+10
        51164.8    1.07962e+11
        53736.9    1.22272e+11
        56437.6    1.38397e+11
        59273.4    1.56582e+11
          62251    1.77296e+11
        65377.5    2.00726e+11
        68660.2    2.27171e+11
        72107.1    2.56942e+11
        75726.4    2.90783e+11
        79526.6    3.29108e+11
        83516.9    3.72439e+11
        87706.6     4.2114e+11
        92105.8    4.76384e+11
          96725     5.3903e+11
         101575    6.09961e+11
         106668    6.89571e+11
         112015    7.79714e+11
         117630    8.82053e+11
         123525    9.98059e+11
         129715    1.12812e+12
         136215    1.27516e+12
         143040    1.44227e+12
         150206    1.63185e+12
         157730    1.84422e+12
         165630    2.08402e+12
         173925    2.35681e+12
         182636     2.6664e+12
         191781      3.013e+12
         201384    3.40401e+12
         211467    3.84921e+12
         222055    4.35451e+12
         233171    4.91991e+12
         244844    5.55743e+12
         257100    6.28388e+12
         269969    7.10821e+12
         283481    8.03019e+12
         297669    9.06952e+12
         312566    1.02547e+13
         328209     1.1599e+13
         344633     1.3102e+13
         361879    1.47962e+13
         379986    1.67296e+13
         399000    1.89212e+13
         418964    2.13706e+13
         439926    2.41324e+13
         461936    2.72858e+13
         485046    3.08578e+13
         509313    3.48489e+13
         534792    3.93505e+13
         561546    4.44934e+13
         589637    5.03142e+13
         619133    5.68159e+13
         650103    6.41536e+13
         682622    7.25403e+13
         716767    8.20236e+13
         752620    9.26142e+13
         790264    1.04574e+14
         829792     1.1825e+14
         871295    1.33698e+14
         914874    1.50946e+14
         960631    1.70418e+14
    1.00868e+06    1.92647e+14
    1.05912e+06    2.17698e+14
    1.11209e+06    2.45654e+14
    1.16771e+06     2.7739e+14
    1.22611e+06    3.13925e+14
    1.28743e+06    3.55282e+14
    1.35182e+06     4.0151e+14
    1.41942e+06    4.53217e+14
    1.49041e+06    5.11104e+14
    1.56494e+06    5.75398e+14
     1.6432e+06    6.46402e+14
\end{verbatim}
\end{tiny}

\clearpage

\subsubsection*{imactint.dat}
\begin{tiny}
\verbatiminput{imactint.dat}
\end{tiny}

